\documentclass[conference]{IEEEtran}
\IEEEoverridecommandlockouts

\usepackage{cite}
\usepackage{graphicx}
\usepackage{textcomp}
\usepackage{xcolor}

\usepackage[english]{babel}

\usepackage{float}
\usepackage{subfigure}
\usepackage{tikz}

\usepackage{amsmath,amssymb,amsfonts}
\usepackage{algorithmic}
\usepackage[ruled,linesnumbered]{algorithm2e}
\usepackage{soul}
\usepackage{wrapfig}

\usepackage{url}

\def\DONE{{\color{red}DONE }}

\newcommand{\DEL}[1]{\iffalse #1 \fi}

\usepackage{ifthenx}
\newtheorem{ins}{\textbf{\textit{O\hspace{-2.8pt}}}}
\newcommand{\insref}[1]{\textbf{\textit{O\hspace{-2.6pt}}}~\ref{#1}}

\usepackage{cleveref}
\crefname{section}{\S}{\SS}
\crefformat{section}{\S#2#1#3} 
\crefformat{subsection}{\S#2#1#3}
\crefformat{subsubsection}{\S#2#1#3}

\newcommand*\circled[1]{\tikz[baseline=(char.base)]{
    \node[shape=circle,draw,inner sep=0.8pt] (char) {#1};}}

\makeatletter
\newcommand{\observation}[2]{%
  \phantomsection
  #1\def\@currentlabel{\unexpanded{#1}}\label{#2}%
}
\makeatother

\newcommand{\squishlist}{
\begin{list}{$\bullet$}
  { \setlength{\itemsep}{0pt}
     \setlength{\parsep}{0pt}
     \setlength{\topsep}{0pt}
     \setlength{\partopsep}{0pt}
     \setlength{\leftmargin}{0em}
     \setlength{\labelwidth}{0em}
     \setlength{\labelsep}{0.2em} } }

\newcommand{\squishlisttwo}{
\begin{list}{$\bullet$}
  { \setlength{\itemsep}{0pt}
     \setlength{\parsep}{0pt}
    \setlength{\topsep}{0pt}
    \setlength{\partopsep}{0pt}
    \setlength{\leftmargin}{2em}
    \setlength{\labelwidth}{1.5em}
    \setlength{\labelsep}{0.5em} } }

\newcommand{\squishend}{
  \end{list}  }

\begin{document}

\title{Straggler Tolerant and Resilient DL Training on Homogeneous GPUs}

\author{
Zeyu Zhang, Haiying Shen\\
University of Virginia, USA\\
Email: qxc4fh@virginia.edu, hs6ms@virginia.edu
}


\maketitle



\begin{abstract}
Despite the popularity of homogeneous GPU-based deep learning (DL) training, the prevalence, causes and impact of stragglers and the effectiveness of existing straggler mitigation approaches are still not well understood in this scenario due to limited research on these questions. To fill this gap, we conducted comprehensive experiments and found that stragglers remain widespread due to CPU and bandwidth usage imbalances. Additionally, existing mitigation methods that switch from synchronous stochastic gradient descent (SSGD) to asynchronous SGD (ASGD) may not improve Time-To-Accuracy (TTA) and can even generate more stragglers due to its higher resource consumption. To address these newly found problems, we propose the Straggler Tolerant And Resilient DL training system (STAR). STAR includes new synchronization modes that group workers for each parameter updating. It has a heuristic and an ML method to choose the optimal synchronization mode for minimizing TTA, and reallocates resources to support the selected mode while minimizing the impact on co-located jobs. Moreover, it proactively prevents stragglers by avoiding overloading the CPU and bandwidth resources. 
Our trace-driven evaluation on AWS shows that STAR generates 48-84\% and 51-70\% lower TTA than state-of-the-art systems in the PS and all-reduce architectures, respectively, while maintaining the converged accuracy of SSGD. The code for STAR is open-sourced.

\end{abstract}

\section{Introduction}\label{sec:Intro}

Deep learning (DL) techniques, employed in diverse domains like computer vision \cite{he2016resnet} and natural language processing \cite{kenton2019bert}, face challenges due to the escalating scales of training datasets and models. This growth results in increasingly time-consuming, resource-intensive, and costly DL training processes. For instance, the training of ChatGPT-3.5 consumed \$4.6M and took 34 days using 1023 A100 GPUs \cite{chatgpt_statistics,openai_gpt3}.  Training ChatGPT-4 requires 10 times more data samples than ChatGPT-3.5. Therefore, it is crucial to reduce the time, resources, and cost.
The parameter server (PS) \cite{li2014ps} and all-reduce (AR)~\cite{ringallreduce} architectures are extensively employed in distributed DL training. In SSGD \cite{gerbessiotis1994direct}, each PS updates parameters only when it receives gradients from all workers, making it susceptible to stragglers if a worker is slow. ASGD \cite{dean2012large} addresses stragglers by allowing each PS to update parameters upon receiving gradients from any worker. However, this approach may lead to lower \emph{accuracy improvement} (i.e., accuracy increased in a certain time period or a certain number of steps), as it incorporates stale gradients from stragglers \cite{oyama2016predicting, meng2019convergence, zhang2016staleness}. As a result, stragglers in both SSGD and ASGD increase training time, resource consumption and cost.\looseness=-1

Previous research has predominantly addressed straggler issues in DL training on CPUs~\cite{dean2012large, harlap2016flexrr, harlap2017proteus, qiao2018litz, chen2020lbbsp} or heterogeneous GPUs \cite{chen2015mxnet, lian2018asynchronous, zhao2019dynamic, luo2019hop, zhou2019falcon, chen2019round, geng2019accelerating, chen2020lbbsp, zeno++}, where stragglers are evident. Recent studies \cite{understanding-straggler,megascale,malleus,adaptra,xu2021lgc,li2021sync} highlight that even in homogeneous GPU clusters, DL training jobs can still exhibit straggler phenomena. Training on homogeneous GPUs is common, introducing new questions about the prevalence, specific causes, and impact of stragglers in this context, as well as the effectiveness of existing straggler mitigation approaches.
Unfortunately, limited research has been conducted to address these questions.



To address these inquiries, we conducted an extensive trace-driven measurement study, yielding valuable insights. Our investigation unveiled the persistence of stragglers in this scenario, attributed to imbalances in CPU and bandwidth usage, resulting in significant (32\%) training time delays and the GPU resource wastage (24\%). Additionally, our study indicates that switching from SSGD to ASGD may not enhance TTA and could potentially generate more stragglers due to its elevated CPU and bandwidth consumption. In response to our findings, we propose the Straggler Tolerant And Resilient DL training system (STAR). STAR introduces new synchronization modes, and determines the optimal mode for minimizing TTA in both the PS and AR architectures (with the ring architecture as an example in this paper). The system incorporates the following methods, each applicable to both PS and AR.

\looseness=-1


\squishlist

\item[(1)] \textbf{Straggler prediction.} We found that stragglers' durations exhibit wide variability, ranging from 0.1 to 500 seconds. Consequently, existing approaches~\cite{li2021sync, chen2020lbbsp, harlap2016flexrr, chen2020elastic}, predicting stragglers based on fixed durations (e.g., 5s), lack precision.
    To address this, we propose a straggler prediction method that utilizes time series data of each worker's CPU and bandwidth resources to forecast its iteration time.\looseness=-1

\item[(2)] \textbf{Static and dynamic x-order synchronization modes.}
We observed that the iteration times of a job's workers in an iteration spread in a range and exhibit clustering patterns. Hence, we propose a static-$x$-order synchronization mode, which uses the gradients from $x~(1<x<N)$ workers for each parameter update for a job with $N$ workers, and a dynamic-$x$-order synchronization mode, which uses the gradients from workers with similar predicted iteration times for each update. For the AR architecture, we propose removing $x$ slow workers from the ring and connecting them to high-bandwidth ring workers (parents), which wait for $t_w$ time after computation to collect and aggregate gradients from them. Different values of $x$ and $t_w$ form different synchronization modes for the AR architecture.

\item[(3)] \textbf{Synchronization mode determination.}
We propose a heuristic method (STAR-H) and a machine learning (ML)-based method (STAR-ML) to estimate the TTA performance of different synchronization modes mentioned above and select the optimal one upon predicting a straggler. The system runs STAR-H first and switches to STAR-ML once the ML model is trained using the data from STAR-H.

\item[(4)] \textbf{Resource-aware straggler prevention.}

\begin{itemize}

\vspace{-0.00in}\item\textbf{Preventing stragglers upon mode change.}  To prevent the selected synchronization mode for a job from generating more stragglers, STAR strategically reallocates resources from co-located jobs to this job within a server node while minimizing the performance impact on them. In the co-located jobs, STAR equalizes iteration times within a worker group (whose reported gradients are for one parameter update or for one aggregation in AR) by depriving resources from faster workers, as this does not affect TTA. Further resource deprivation targets jobs less sensitive to resource decrease and are in later training phases.\looseness=-1

\vspace{-0.00in}\item\textbf{Proactively preventing stragglers.} We found that a PS consumes significantly more CPU and bandwidth resources than a worker. Similarly, a parent in AR consumes more CPU and bandwidth resources. Considering that stragglers can result from both iteration time increases and decreases, STAR prevents stragglers by balancing the number of PSs among servers, and the number of parents among servers in AR. In addition, recognizing diverse~\cite{xu2021lgc,li2021sync} and time-varying~\cite{time-varying} bandwidth capacities among servers, STAR organizes the workers in the PS architecture and the children in AR into a tree for communication, amortizing the communication overhead on the PSs or parents.

\end{itemize}

\squishend

\noindent \textbf{Contribution.} In summary, previous straggler mitigation approaches may overlook the possibility of not improving TTA and even generating more stragglers. In contrast, STAR stands out as the first approach that meticulously selects the optimal synchronization mode, considering both existing and newly proposed modes, while balancing resource consumption to prevent stragglers.

We conducted extensive trace-driven experiments on real testbed built on AWS.
STAR generates 48-84\% and 51-70\% lower TTA than state-of-the-art systems in PS and AR, respectively, while maintaining a similar converged accuracy as SSGD. We distributed our source code in GitHub~\cite{ourcode}.

\section{Preliminaries}


During a DL training job, workers conduct pre-processing data using CPUs. It first moves a mini-batch of examples from disk to CPU memory, converts raw data into PyTorch tensors, and then moves tensors to GPU memory.
Workers use GPUs to compute gradients, which are then sent to the PS (or other workers in AR). The PS (or each worker in AR) updates the parameters, and sends them back to the workers. Data receivers employ busy-polling~\cite{li2020packetio} to monitor incoming data, consuming CPU. Bandwidth is consumed during the transfer of gradients and parameters.
Let $T_{k}$ represent the iteration time of worker $k$ among the $N$ workers of a job. The iteration time deviation and its ratio of an iteration $j$ are calculated as $D_j=\max_{k=1}^{N} T_k- \min_{k=1}^{N} T_k$ and $\frac{D_j}{\min_{k=1}^{N} T_k}$, respectively. Worker $i$'s iteration time deviation ratio is calculated as $d_i=\frac{T_{i}-\min_{k=1}^{N} T_{k}}{\min_{k=1}^{N} T_{k}}$, and it is identified as a straggler if $d_i> 20\%$~\cite{harlap2016flexrr}.

\DEL{For a given DL training job, a job scheduler~\cite{chen2020elastic, aurick2021pollux, gu2022liquid, wang2020job, gu2019tiresias} assigns its PS and each worker to servers. Each worker stores the training dataset in the CPU memory. A mini-batch of samples selected from the dataset is copied to GPU memory for training every iteration. Model parameters are in the GPU memory. A worker uses GPU to compute gradients, which are temporarily copied from GPU memory to CPU memory before being sent to the PS. 
When the PS receives gradients from the worker, it uses the CPU to actively check the message queue (i.e., busy-polling~\cite{li2020packetio}) for incoming data to avoid latency. Workers sending gradients and the PS receiving gradients consume bandwidth.
The PS always stores one copy of parameters and also gradients in its CPU memory. In SSGD, when the PS receives the first gradient copy from a worker, it stores it in its CPU memory. When the next gradient copy from another worker comes, the PS immediately adds it to the previous gradient copy using CPU. After the PS adds all workers' gradient copies, it updates parameters using CPU. In ASGD, every time when the PS receives a gradient copy from a worker, it stores it in its CPU memory and updates parameters for this worker using CPU~\cite{jiang2020unified}.
The PS sends the updated parameters to workers, which consumes bandwidth, and a worker receives the parameters, which consumes CPU and bandwidth.
A worker temporarily stored its received parameters in its CPU memory, and then moves them to its GPU memory. Then, the worker loads another mini-batch from its CPU memory to its GPU memory to start the next iteration.}

\section{Experimental Analysis}\label{sec:exp_measure}


Unless otherwise specified, the experiment settings are as follows. We used 5 AWS EC2 p4d.24xlarge GPU instances and 3 m4.16xlarge CPU instances. Each p4d.24xlarge instance features 8 NVIDIA A100 GPUs, 96 vCPUs, and 1152GB memory, and each m4.16xlarge instance has 64 vCPUs and 256GB memory. We used the Microsoft Philly Trace~\cite{phillytrace} and selected a time interval from October 9 to October 13, 2017, containing 350 jobs. The number of workers of each job was randomly selected between 4 and 12, , with an attempt to place them in the same GPU instance; otherwise, other GPU instances were utilized. The number of PSs of a job was randomly selected between 1 and the number of its workers. Following industry practice, we randomly chose the configuration for running a job's PSs -- either on the job's GPU servers or on separate CPU servers. In the former configuration, if the GPU servers lacked adequate CPU resources, other GPU servers were utilized. Each job selected a model from ten models for image classification on CIFAR-10~\cite{krizhevsky2009cifar} or Natural language processing (NLP) on WikiText-2~\cite{wikitext}. The former models include ResNet20, ResNet56, VGG13, VGG16, DenseNet121, AlexNet, GoogleNet, and MobileNet, and the latter models include LSTM and Transformer.


We set the worker mini-batch size to 128 samples~\cite{li2021sync}, with learning rates of 0.1 for ResNet and 0.01 for other models~\cite{he2016deepresidual,li2021sync,smith2018disciplined}, decaying to 0.01 and 0.001 at the 32k$^{th}$ and 48k$^{th}$ steps to prevent overshooting~\cite{li2021sync,ding2021decay}. All jobs used the PS architecture and terminated upon convergence~\cite{ranjan2016survey}, defined as accuracy or perplexity changing by less than 0.001 over five evaluations spaced 40s apart~\cite{li2021sync}. The target accuracy and perplexity for TTA matched the converged values achieved by ASGD. Resource usage was logged every second.\looseness=-1

\DEL{\textbf{Concurrent-job measurement.} We performed a concurrent workload using the Microsoft Philly Trace~\cite{phillytrace}, which is part of the real-world first-party DNN training workloads from Microsoft's Philly clusters. This trace has 117325 jobs submitted between 08/07/2017-12/22/2017.
We selected an interval from October 9 to October 13, 2017 that contains 350 jobs and used submission time for each job. A job randomly selected a model from the ten models for NLP and image classification to train. Each PS or worker was randomly placed onto one of the instances. 

\textbf{Single-job measurement.} {\color{blue} We trained each model individually in this measurement. Each PS or worker used one instance exclusively.}  ({\color{red} Moved to above}this setting is also needed in "Concurrent-job measurement," why it is here?).  
}


\vspace{-0in}
\subsection{Existence of Stragglers and Causes}\label{observe:existence_of_stragglers}
\vspace{-0in}

\begin{figure}[h]\vspace{-0.0in}
    \centering
    \subfigure[Iteration time.\label{fig:sub:cdf_time_devi_ratio_350_jobs}]
    {\includegraphics[width=0.49\columnwidth,height=2cm]{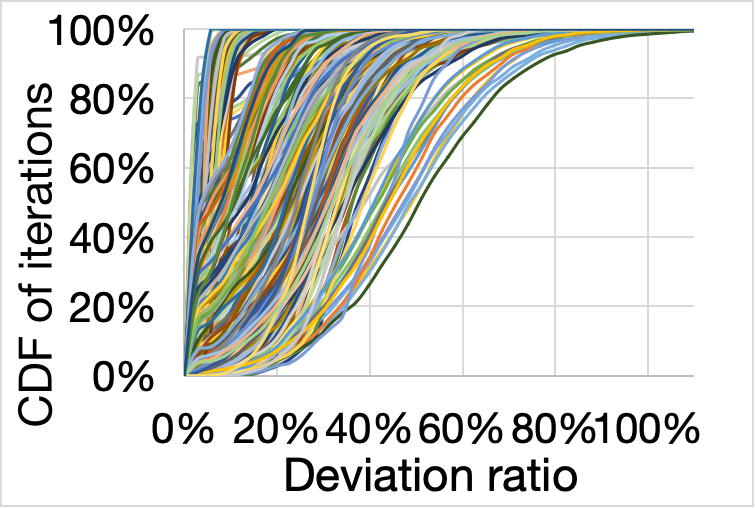}}
    \subfigure[Computation time.\label{fig:sub:cdf_comp_ratio_350_jobs}]
    {\includegraphics[width=0.49\columnwidth,height=2cm]{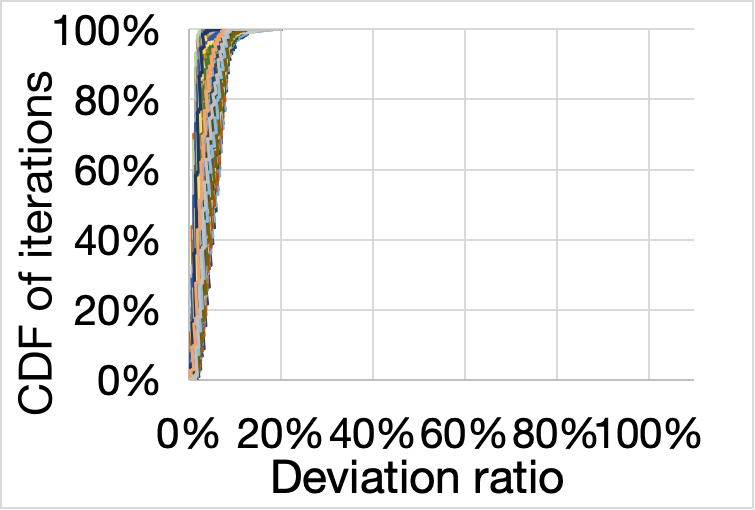}}
    \subfigure[Pre-processing time.\label{fig:sub:cdf_preproc_ratio_350_jobs}]
    {\includegraphics[width=0.49\columnwidth,height=2cm]{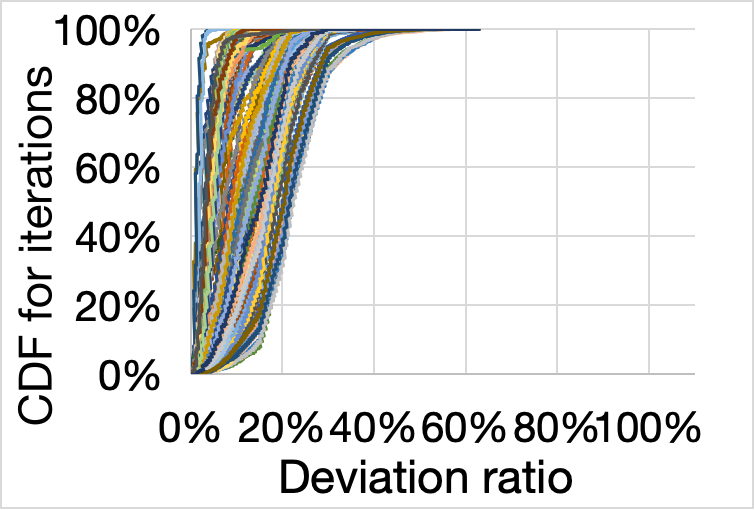}}
    \subfigure[Communication time.\label{fig:sub:cdf_comm_ratio_350_jobs}]
    {\includegraphics[width=0.49\columnwidth,height=2cm]{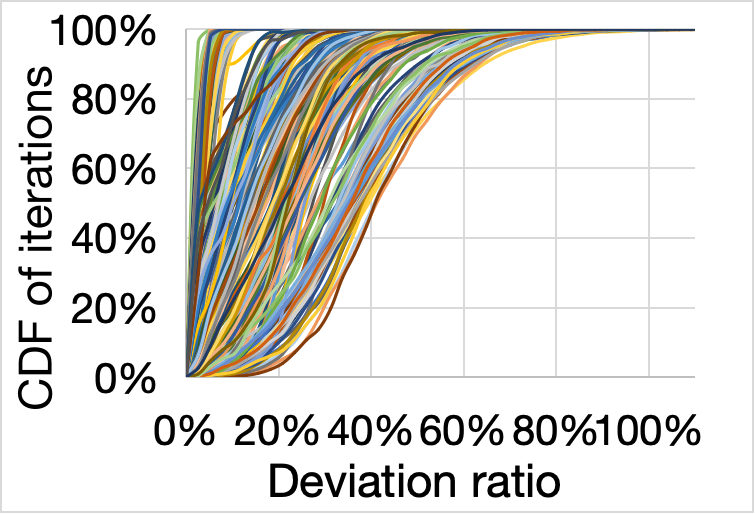}}
    \vspace{-0.15in} \caption{\small{CDF of iterations vs. deviation ratios.}} 
   \vspace{-0in}
    \label{fig:cdf_devi_ratio_350_jobs}
\end{figure}

Fig.~\ref{fig:cdf_devi_ratio_350_jobs} shows the CDF of iterations versus deviation ratios of workers' iteration, GPU computation, pre-processing, and communication times across 350 jobs. Fig.~\ref{fig:sub:cdf_time_devi_ratio_350_jobs} reveals varying iteration-time deviations: all jobs have 0-98\% straggler iterations, and 47\% have over 50\%. Fig.~\ref{fig:sub:cdf_comp_ratio_350_jobs} shows no stragglers from GPU computation. Fig.~\ref{fig:sub:cdf_preproc_ratio_350_jobs} indicates 18\% of jobs have pre-processing stragglers, with 0-70\% affected iterations and 7\% having over 50\%. Fig.~\ref{fig:sub:cdf_comm_ratio_350_jobs} shows 83\% of jobs experience communication stragglers, with 0-97\% affected iterations and 32\% having over 50\%. Overall, stragglers add about 32\% latency and waste 24\% of GPU resources compared to the case where each iteration takes the average worker time $\bar{T}$. Fig.~\ref{fig:ratio_comm_iter} shows communication accounts for 2-93\% of iteration time, with 75\% of ratios in [50\%, 93\%], confirming communication's dominance. Across all iterations, 65\% experience stragglers.\looseness=-1

%
%

\begin{figure}[h]\vspace{-0.05in}
    \centering
    \begin{minipage}[t]{0.235\textwidth}
  \centering
\includegraphics[width=0.95\columnwidth,height=2cm]{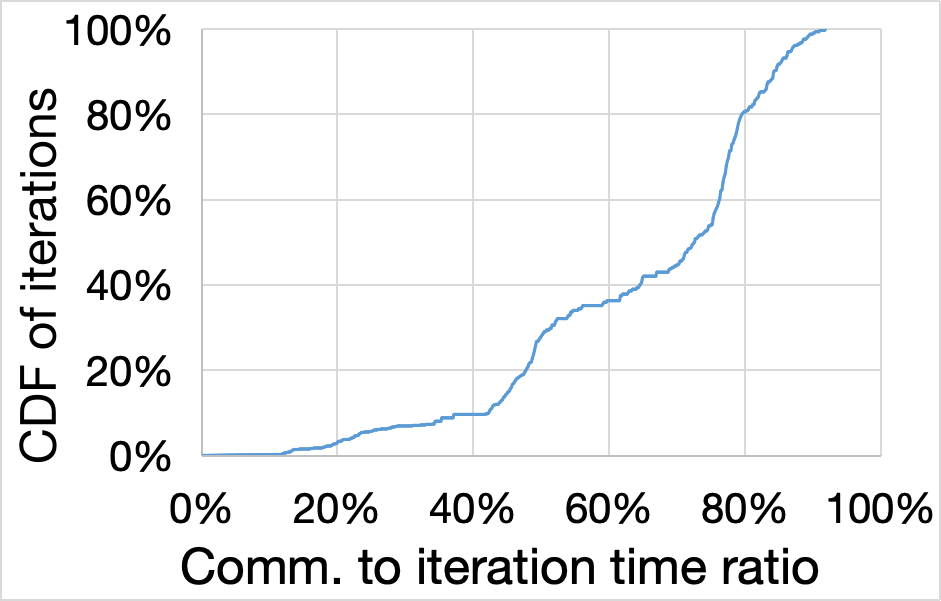}\label{fig:ratio_comm_iter}
\vspace{-0.15in}\caption{\small{Communication time ratio.}}
\label{fig:ratio_comm_iter}\vspace{-0.1in}
    \end{minipage}%
\begin{minipage}[t]{0.235\textwidth}
  \centering
\includegraphics[width=0.95\columnwidth,height=2cm]{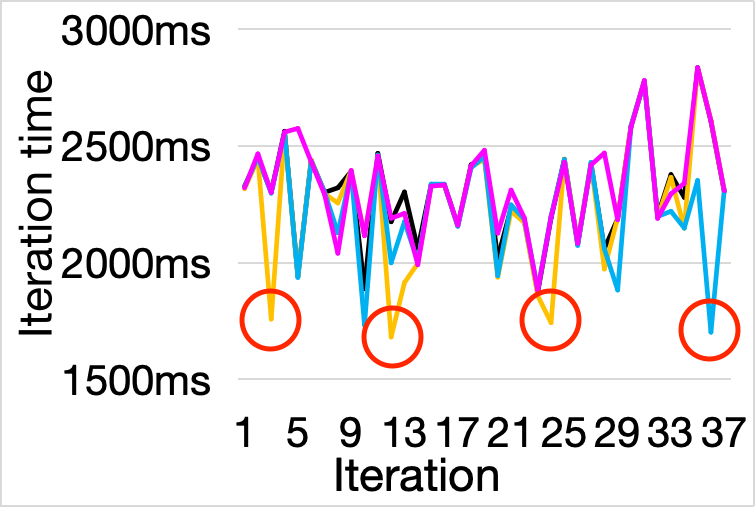}\label{fig:sub:wrk_iter_time_densenet121-1}
\vspace{-0.15in}\caption{\small{Iteration time.}}
\label{fig:sub:wrk_iter_time_densenet121-1}\vspace{-0.2in}
\end{minipage}\vspace{-0.15in}
\end{figure}

\looseness=-1


\DEL{\begin{figure}[H]
    \centering
    \subfigure[Iteration time.\label{fig:sub:cdf_time_devi_ratio_350_jobs_line}]
    {\includegraphics[width=0.32\columnwidth,height=1.5cm]{figures/cdf_time_devi_ratio_350_jobs_line.pdf}}
    \subfigure[Computation time.\label{fig:sub:cdf_comp_ratio_350_jobs_line}]
    {\includegraphics[width=0.32\columnwidth,height=1.5cm]{figures/cdf_comp_ratio_350_jobs_line.pdf}}
    \subfigure[Communication time.\label{fig:sub:cdf_comm_ratio_350_jobs_line}]
    {\includegraphics[width=0.32\columnwidth,height=1.5cm]{figures/cdf_comm_ratio_350_jobs_line.pdf}}
    \caption{CDF of deviation ratios of workers of all jobs.}
    \label{fig:cdf_devi_ratio_350_jobs_line}
\end{figure}}


%

\begin{figure}[h]\vspace{-0.1in}
  \centering
\includegraphics[width=1\columnwidth,height=2.5cm]{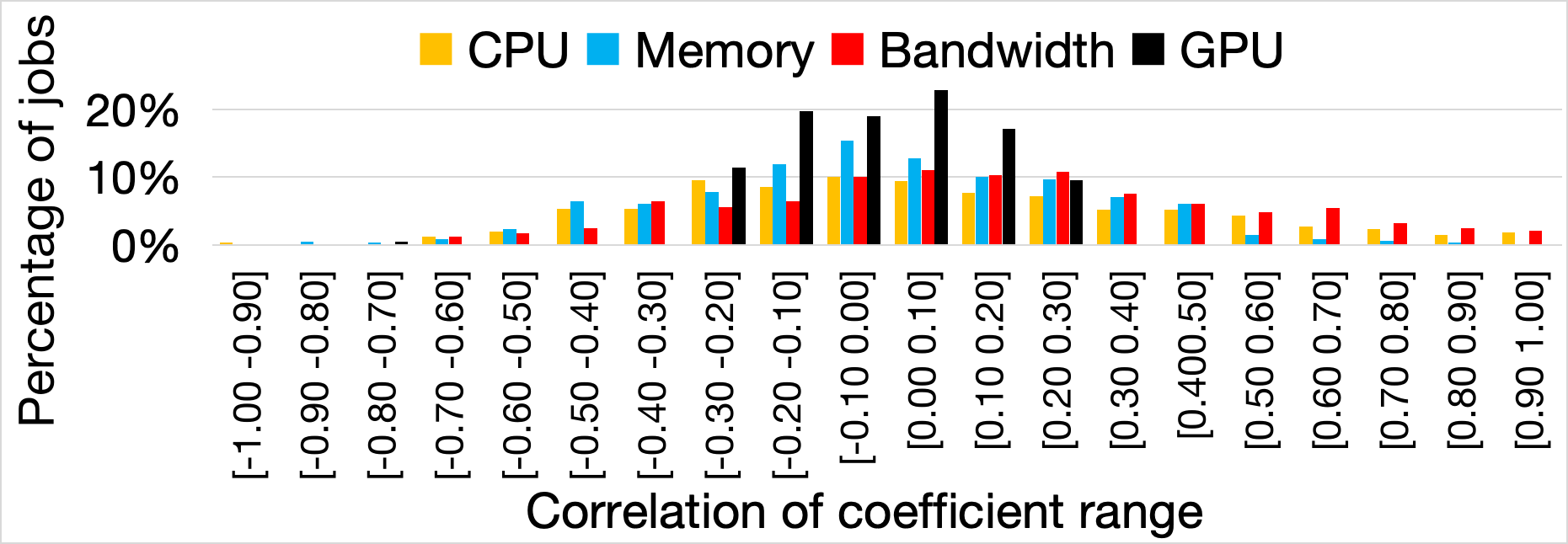}
\vspace{-0.3in}\caption{\small{Correlation coefficients.}}
\label{fig:corrcoef_cpu_bw_peak_and_20iter_around}
\vspace{-0.15in}
\end{figure}

For each job and resource type, we computed the correlation between the max-min difference in resource usage and iteration time across all iterations. Fig.~\ref{fig:corrcoef_cpu_bw_peak_and_20iter_around} shows the correlation distributions for 350 jobs by resource type. GPU correlations fall within [-0.3, 0.3], while 13.8\% of CPU, 17.1\% of bandwidth, and 3.2\% of memory coefficients lie in [0.5, 1]. Server memory usage ranges from 2\% to 16\%. These results indicate CPU and bandwidth as key straggler contributors. Alibaba Group's DL job trace analysis~\cite{alibaba-trace} further confirms that workers can become stragglers during training.\looseness=-1



\DEL{\vspace{-0.1in}
\begin{ins}~\label{ins_straggler}
In homogeneous GPU based DL training, stragglers commonly exist caused by CPU and bandwidth usage, and they could delay 32\% training time and waste 24\% GPU resource. \looseness=-1 
\end{ins}\vspace{-0.1in}
}

\begin{figure}[h]\vspace{-0.05in}
    \centering
\begin{minipage}[b]{0.235\textwidth}
\includegraphics[width=0.985\columnwidth,height=2cm]{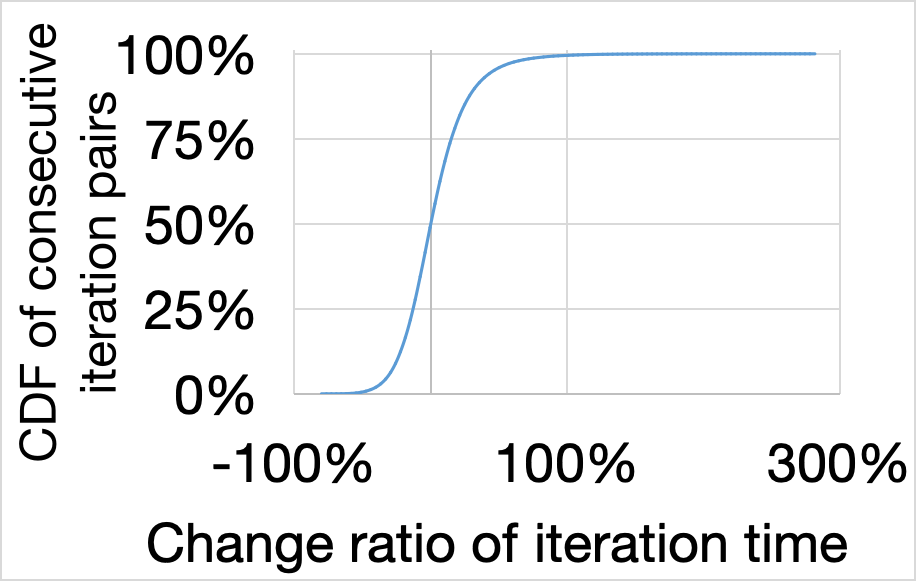}
\vspace{-0.3in}\caption{Iteration time change of a worker. }
\label{fig:dynamicChanging}\vspace{-0.05in}
\end{minipage}%
\begin{minipage}[b]{0.235\textwidth}
    {\includegraphics[width=0.985\columnwidth,height=2cm]{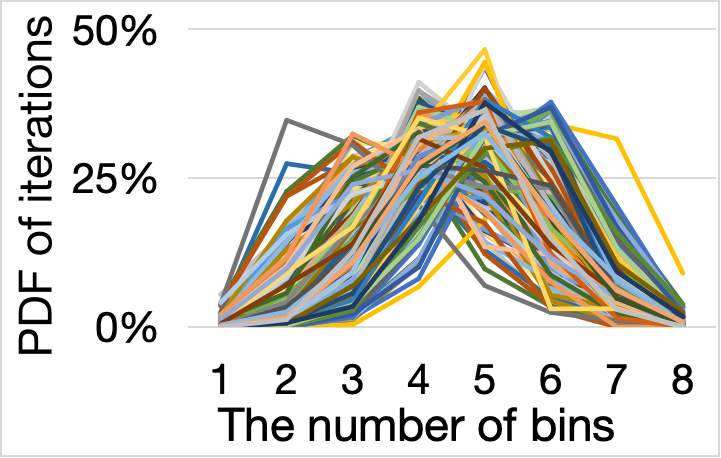}}
 \vspace{-0.3in}   \caption{Num. of bins for workers' iteration times.}
    \label{subfig:wrk_iter_clustering_each}\vspace{-0.05in}
\end{minipage}
\end{figure}

\begin{figure*}[h]
\centering
\begin{minipage}[b]{0.25\textwidth}
 {\includegraphics[width=0.98\columnwidth,height=2cm]{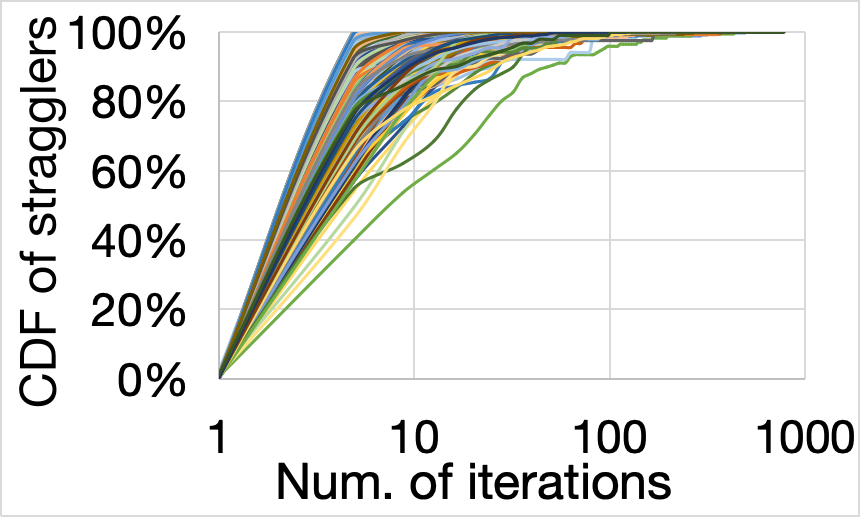}}
   \vspace{-0.08in}  \caption{Num. of iterations a straggler lasts.}
    \label{fig:strg_last_for_iter} \vspace{-0.065in}
    \end{minipage}%
\begin{minipage}[b]{0.75\textwidth}
\subfigure[PS's CPU usage.\label{subfig:avg_cpu_ps}]{\includegraphics[width=0.245\linewidth,height=2cm]{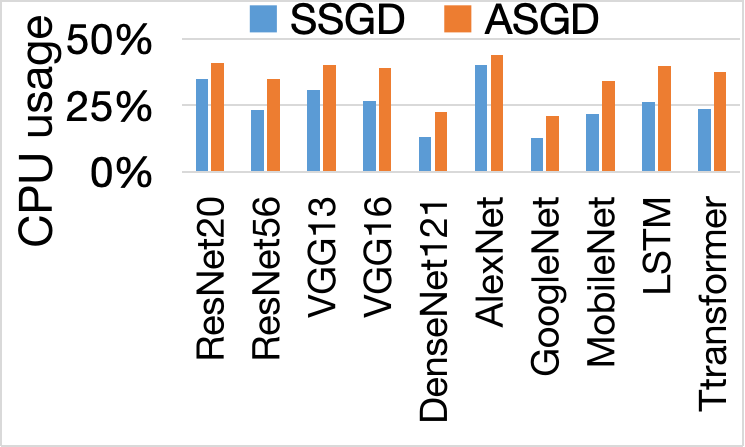}}
\subfigure[Worker1's CPU usage.\label{subfig:avg_cpu_wrk1}]{\includegraphics[width=0.245\linewidth,height=2cm]{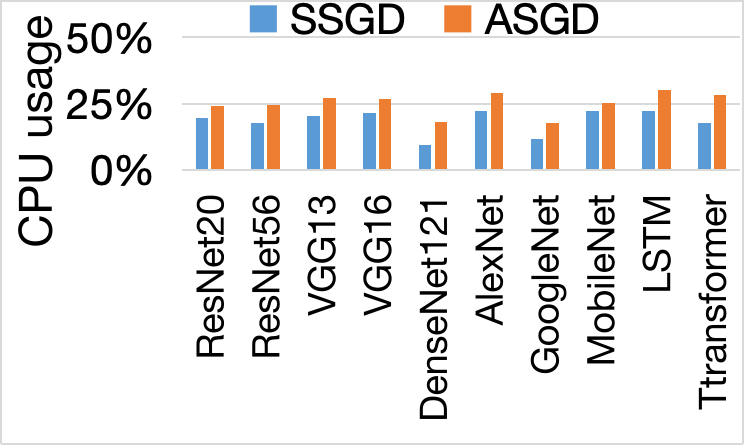}}
\subfigure[PS's BW usage.\label{subfig:avg_bw_ps}]{\includegraphics[width=0.245\linewidth,height=2cm]{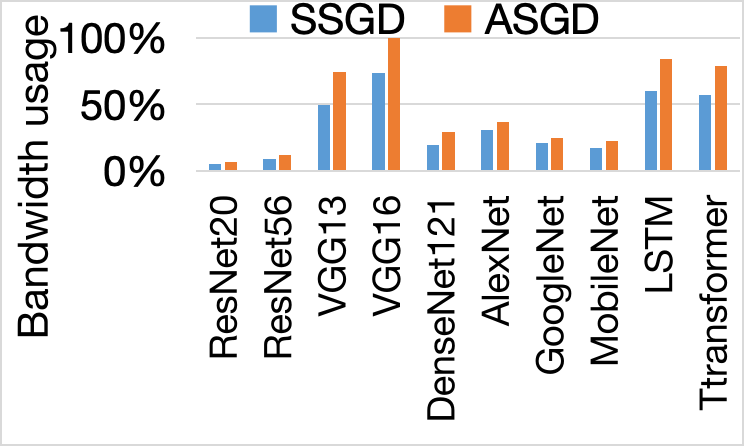}}
\subfigure[Worker1's BW usage.\label{subfig:avg_bw_wrk1}]{\includegraphics[width=0.245\linewidth,height=2cm]{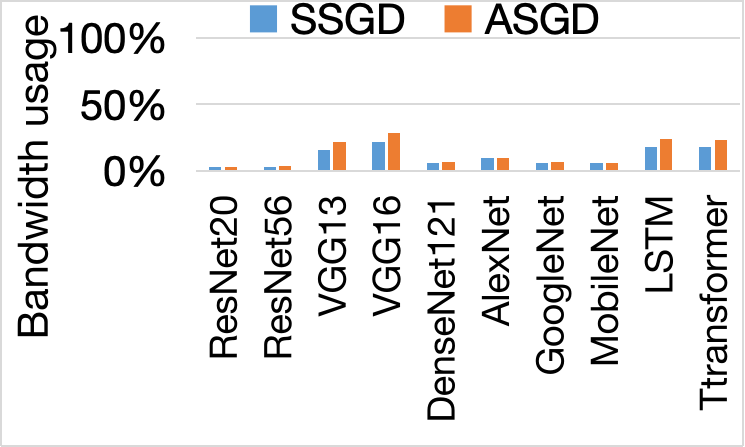}}
\vspace{-0.2in}\caption{Average resource usage of PS and worker1.}\vspace{-0.1in}
\label{fig:ps_worker_aver_load}
\end{minipage}
\vspace{-0.3in}
\end{figure*}

\begin{figure}[h]
    \centering
    \subfigure[CPU usage.\label{subfig:cdf_cpu_ps_num}]{\includegraphics[width=0.49\columnwidth,height=2cm]{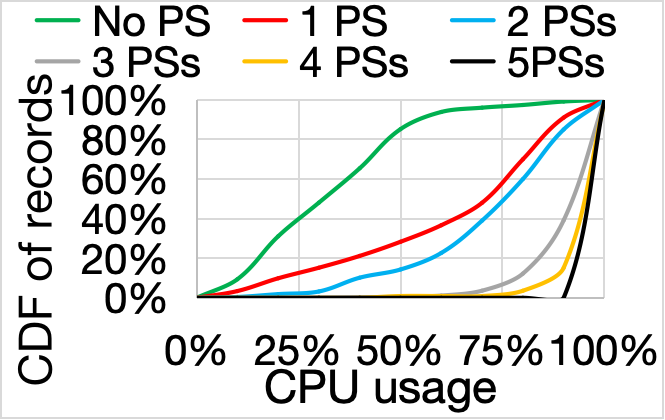}}
    \subfigure[Bandwidth usage.\label{subfig:cdf_bw_ps_num}]{\includegraphics[width=0.49\columnwidth,height=2cm]{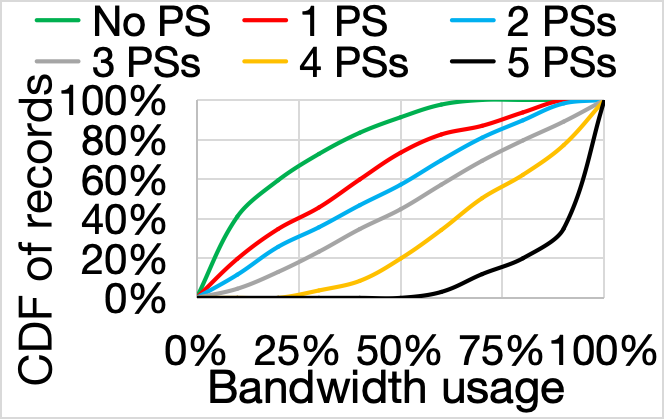}}
\vspace{-0.2in}
        \caption{\small{Servers hosting more PSs consume more resources.}}
    \label{fig:cdf_res_ps_wrk_num}\vspace{-0.25in}
\end{figure}

Fig.~\ref{fig:sub:wrk_iter_time_densenet121-1} shows iteration times of four workers across iterations in DenseNet121. Worker iteration times fluctuate, and large deviations arise from both increases and decreases. Similar patterns are observed for other models (omitted for brevity).\looseness=-1

Fig.~\ref{fig:dynamicChanging} presents the CDF of consecutive iteration pairs versus the iteration time change ratio $\frac{T^{k+1} - T^k}{T^k}$, where $T^k$ is the iteration-$k$ time of a worker. About 23\% and 21\% of pairs show over 20\% increases and decreases, respectively, demonstrating the dynamic variation in worker iteration times.\looseness=-1

\vspace{-0in}
\begin{ins}~\label{ins_straggler} 
In homogeneous GPU based DL training, stragglers commonly exist caused by CPU and bandwidth usage, and they could delay 32\% training time and waste 24\% GPU resource. Straggling occurs due to iteration time increases and decreases.\looseness=-1

\end{ins}\vspace{-0in}

We next analyze the distribution of worker iteration times within each job. The range from 0 to the job's maximum iteration time is divided into eight bins. Fig.~\ref{subfig:wrk_iter_clustering_each} plots the PDF of iterations versus the number of bins containing worker iteration times across 350 jobs. Jobs have 11-42\%, 10-48\%, 4-39\%, 1-32\%, and 0.5-9\% of iterations spanning 4, 5, 6, 7, and 8 bins, respectively.\looseness=-1

\DEL{The results suggest that the worker iteration times in an iteration tend to spread in a range and show a clustering pattern.}


\DEL{
\begin{figure}[H]
\centering
\includegraphics[width=0.6\columnwidth,height=2.5cm]{figures/asgd_x_by_x_time.pdf}
\caption{Accuracy over time for $x$-order updating.}
\label{fig:asgd_x_by_x_time}
\end{figure}
}


\vspace{-0in}
\begin{ins}~\label{ins_x_by_x}
The iteration times of a job's workers in an iteration exhibit clustering patterns within a certain range, and a worker's iteration time varies across different iterations.
\end{ins}\vspace{-0in}

\vspace{-0in}
\subsection{Straggler Prediction}
\vspace{-0in}
Fig.~\ref{fig:strg_last_for_iter} shows the CDF of stragglers versus the number of iterations they persist across 350 jobs. Specifically, 16\% of jobs have stragglers lasting no more than 10 iterations, 37\% have 2-19\% stragglers lasting 10-50 iterations, and 12\% have 1-5\% lasting over 100 iterations. The corresponding lasting times range from 0.1 to 419s. Categorizing stragglers solely by exceeding a fixed duration (e.g., 5s~\cite{li2021sync}) is imprecise, yielding 10.2-22.8\% false positives (FPs) and 4.3-24.8\% false negatives (FNs). Using an LSTM~\cite{lstm} to predict the next-iteration deviation ratio from 100 past values reduces precision only slightly, with 8.7-27.6\% FPs and 25-42.1\% FNs. These results are summarized in Fig.~\ref{fig:relwrk_false_posi_nega} of~\cref{sec:exp}.\looseness=-1

\begin{figure*}[h] 
\centering
\begin{minipage}[b]{0.24\textwidth}
\includegraphics[width=0.95\columnwidth,height=2cm]{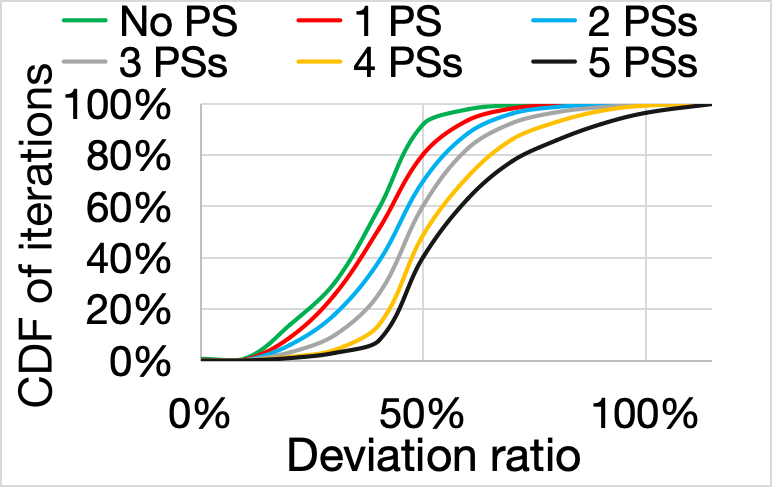}
 \vspace{-0.12in}\caption{Higher deviation ratios in servers with more hosted PSs.}\vspace{0.065in}
\label{fig:cdf_iter_ratio_for_ps_num}
\end{minipage}%
\begin{minipage}[b]{0.76\textwidth}
      \subfigure[Resource usage.\label{subfig:switch_srv_res}]{\includegraphics[width=0.32\linewidth,height=2cm]{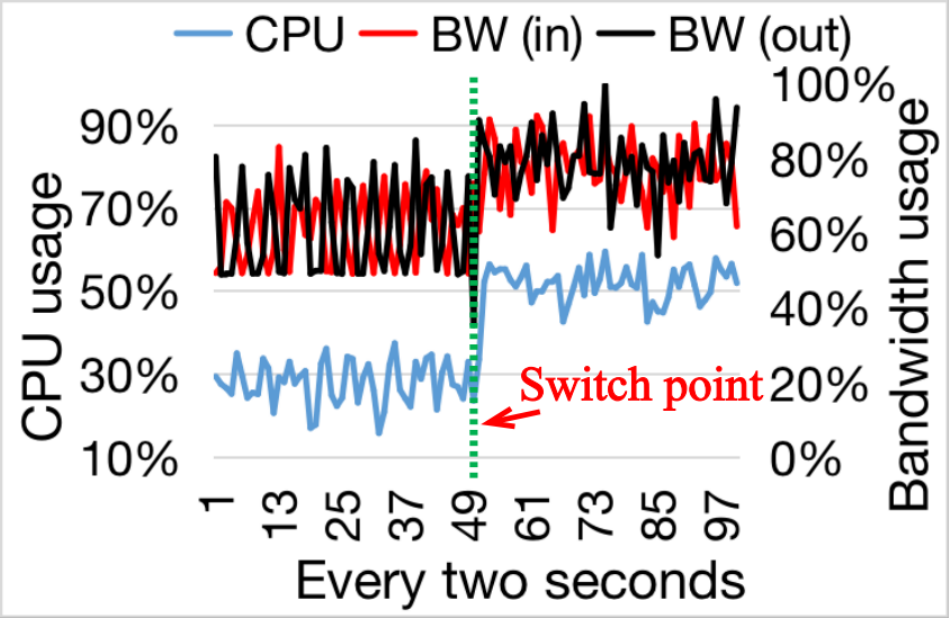}}
    \subfigure[Iteration time of colocated workers.\label{subfig:col_wrk_iter_time}]{\includegraphics[width=0.32\linewidth,height=2cm]{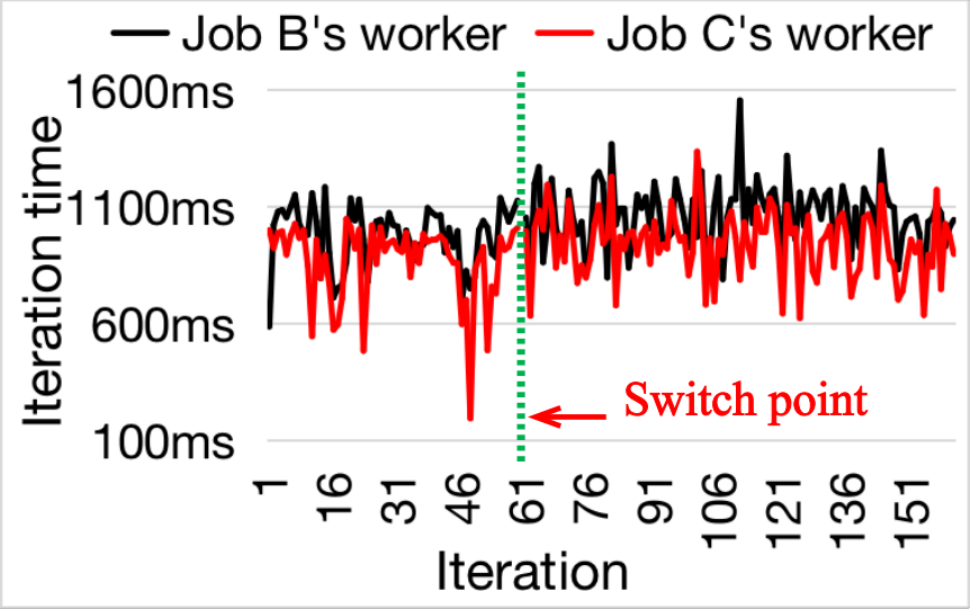}}
    \subfigure[Num. of stragglers.\label{subfig:switch_strg_num}]{\includegraphics[width=0.32\linewidth,height=2cm]{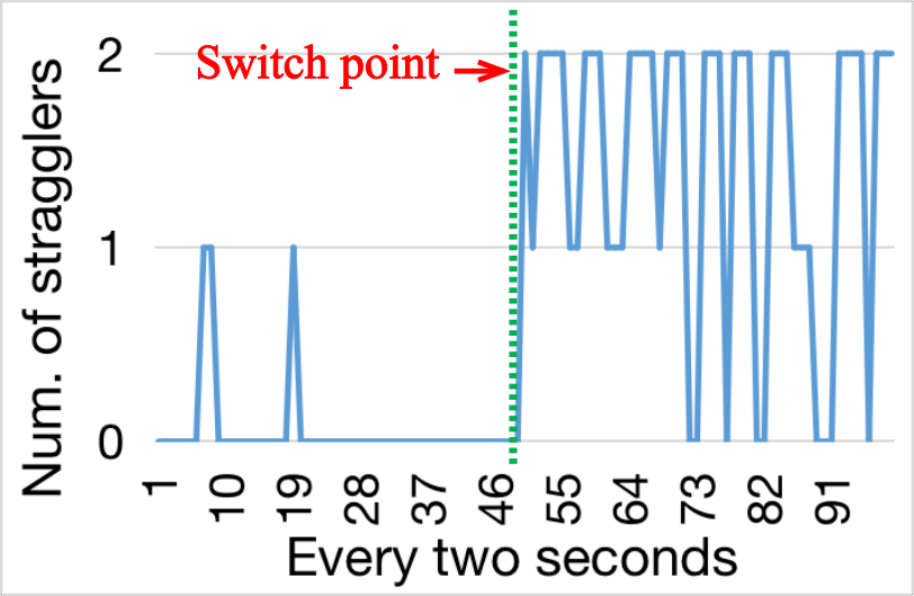}}
 \vspace{-0.12in}   \caption{Switching to ASGD can generate more stragglers.}\vspace{-0.1in}
    \label{fig:switch_strg_num}
  \end{minipage}
\end{figure*}

\begin{figure*}[h]
\centering
\subfigure[No throttling.\label{subfig:tta_cpu_100}]
{\includegraphics[width=0.495\columnwidth,height=2cm]{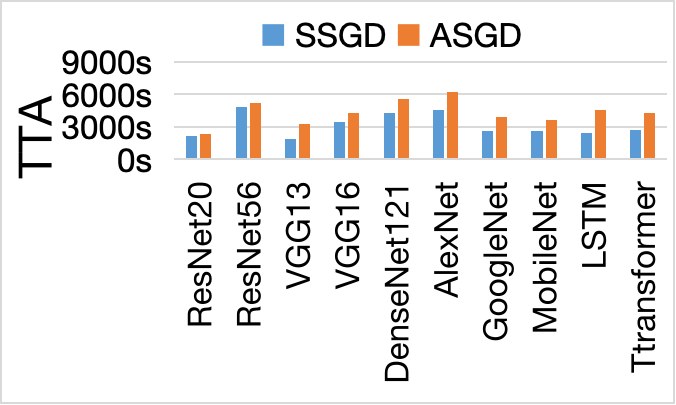}}
\subfigure[Throttling CPU to 75\%.\label{subfig:tta_cpu_75}]
{\includegraphics[width=0.495\columnwidth,height=2cm]{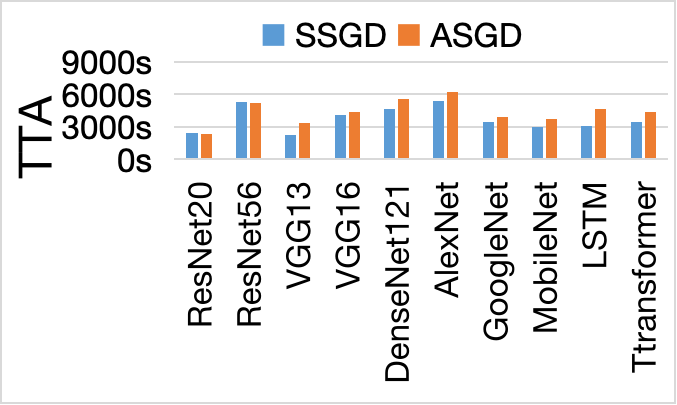}}
\subfigure[Throttling CPU to 10\%.\label{subfig:tta_cpu_10}]
{\includegraphics[width=0.495\columnwidth,height=2cm]{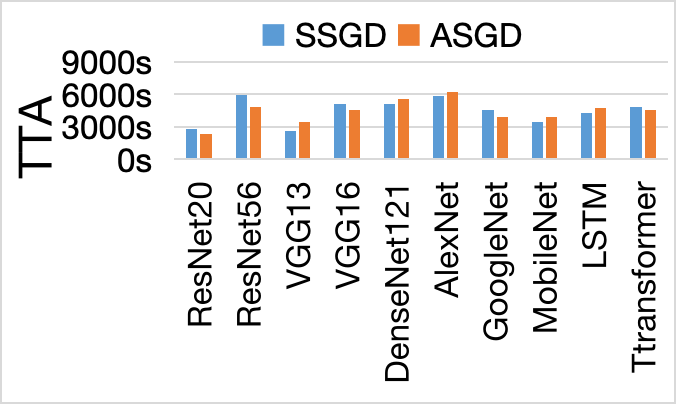}}
\subfigure[Throttling CPU to 5\%.\label{subfig:tta_cpu_5}]
{\includegraphics[width=0.495\columnwidth,height=2cm]{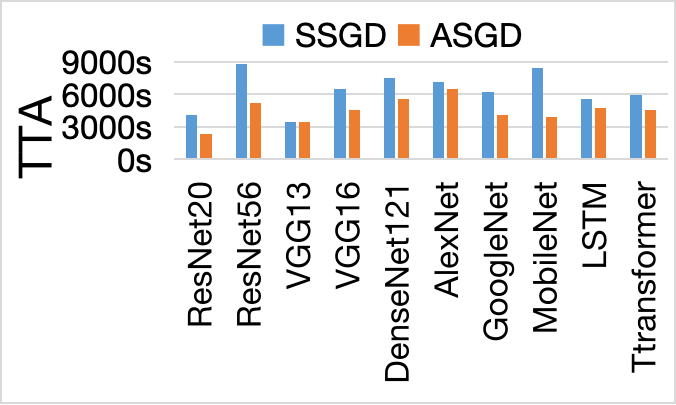}}
\vspace{-0.13in}\caption{Average resource usage of PS and worker1.} 
\label{fig:jct_worker_cpu_strategy}\vspace{-0in}
\end{figure*}

\begin{figure*}[h]
\centering \vspace{-0.1in}
\subfigure[No throttling.\label{subfig:tta_bw_100}]{\includegraphics[width=0.245\linewidth,height=2cm]{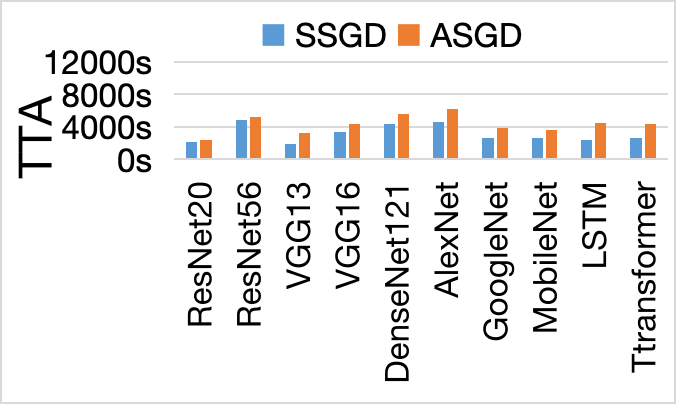}}
\subfigure[Throttling BW to 75\%.\label{subfig:tta_bw_100}]{\includegraphics[width=0.245\linewidth,height=2cm]{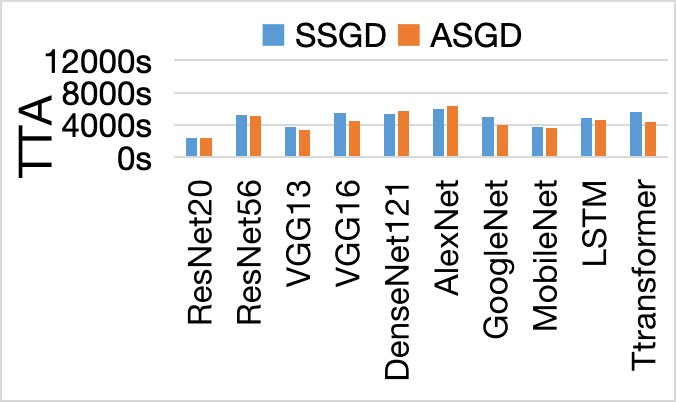}}
\subfigure[Throttling BW to 25\%.\label{subfig:tta_bw_25}]{\includegraphics[width=0.245\linewidth,height=2cm]{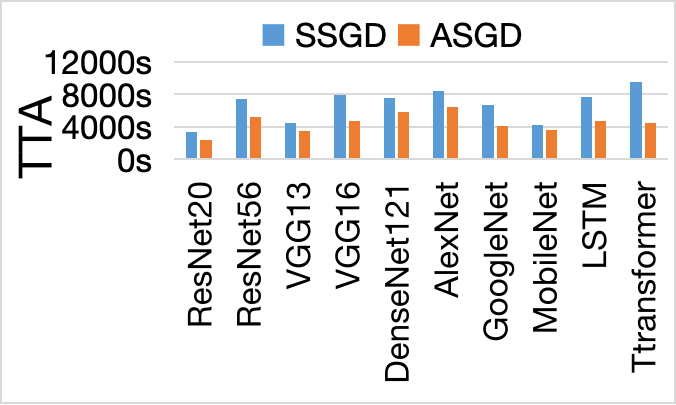}}
\subfigure[Throttling BW to 10\%.\label{subfig:tta_bw_10}]{\includegraphics[width=0.245\linewidth,height=2cm]{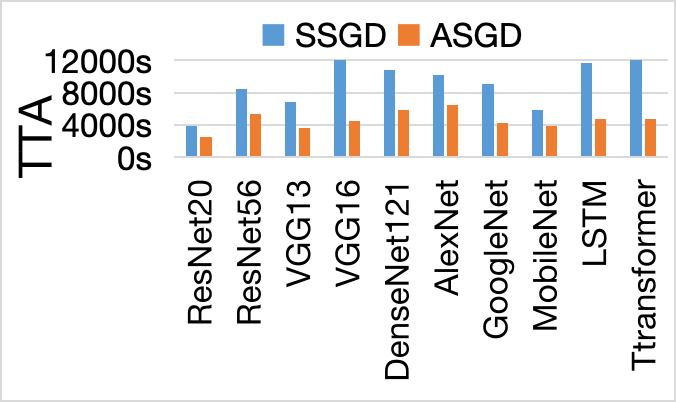}}
\vspace{-0.23in}\caption{TTA when throttling the bandwidth of worker1 in SSGD and ASGD.}\vspace{-0.2in}
\label{fig:jct_worker_bw_strategy1}
\end{figure*}


\vspace{-0in}

\begin{ins}~\label{ins_predict_ratio} Classifying a worker that straggles for a fixed time period to be a straggler, and using ML to predict the future deviation ratio based on past ratios are not sufficiently accurate.\vspace{-0.1in}

\end{ins}\vspace{-0in}

\DEL{In addition, as the CPU throttling increases, different jobs have different TTA increase ratios in SSGD due to different model types as shown in Table~\ref{tab:model_size}. 
For example, the ratio for a small model ResNet56 is 0.13 and 0.78; the ratio for a large model Transformer is 0.81 and 1.97. 
A large model has large parameters. When worker1's CPU is throttled, a large model's worker1 will take more time to receive parameters from the PS than a small model, thus increasing TTA more under higher straggling degree in SSGD.}



\DEL{When the throttle ratio is 10\% and 5\% of its original capacity, in SSGD, the jobs' TTAs are increased by 11\%-103\% and 78\%-300\%, respectively, and in ASGD, the jobs' TTAs are increased by 0\%-6\% and 0\%-7\%, respectively. Therefore,}


\DEL{
\begin{table*}[t]
\centering
\begin{tabular}{| c | c | c | c | c | c | c | c | c | c |}
    \hline
    ResNet20 & ResNet56 & AlexNet & VGG13 & VGG16 & GoogleNet & DenseNet121 & MobileNet & LSTM & Transformer \\
    \hline
    1.02MB & 3.25MB & 9.4MB & 35.9MB & 56.1MB & 23.52MB & 26.5MB & 8.76MB & 46.48MB & 45.8MB \\
    \hline
\end{tabular}
\caption{Parameter size of each model.}
\label{tab:model_size}
\end{table*}
}


\DEL{When the throttle ratio is 25\% and 10\%, the jobs' TTAs are increased by 23\%-311\% and 96\%-753\% in SSGD; and by 0\%-10\% and 4\%-16\% in ASGD. Bandwidth throttling increases TTA in SSGD but barely affects TTA in ASGD.
As the bandwidth throttling increases, different jobs have different TTA increase ratios in SSGD, e.g., it is 0.3 and 1.04 for small model ResNet56 and 0.81 and 1.97 for large model Transformer.
When the throttle ratio is 25\%, some jobs (ResNet56, VGG16, GoogleNet, LSTM, and Transformer) have 14\%-31\% higher TTA in SSGD than in ASGD; other jobs have 4\%-25\% lower TTA in SSGD than in ASGD. When the throttle ratio is 10\%, all jobs have 33\%-72\% higher TTA in SSGD than in ASGD. When the throttle ratio is 25\% or 10\%, similar-size models have different model difference ratios, e.g., it is 2.55 for VGG16 and is 1.98 for Transformer when the throttled ratio 10\%. 
}



\vspace{-0in}
\subsection{Resource Consumption}\label{observe:resource_consumption}
\vspace{-0in}




To analyze the impact of limited CPU and bandwidth in SSGD and ASGD, we measured single-job performance with 4 workers on one server and the PS on another. Fig.~\ref{fig:ps_worker_aver_load} shows the average CPU and bandwidth usage per second for the PS and worker1 in both systems. In ASGD, the PS and worker1 use 11-75\% and 14-97\% more CPU, and 6-29\% and 7-26\% more bandwidth than in SSGD, respectively, since ASGD avoids waiting for slow workers. The PS consumes 5-76\% and 19-87\% higher CPU and 101-250\% and 135-296\% higher bandwidth than worker1 in SSGD and ASGD, respectively, due to parameter updates and busy-polling. Worker1, by contrast, spends CPU on pre-processing and busy-pulling parameters. Resource usage varies by model type and size. Additional results (omitted for space) show that total CPU and bandwidth differences grow with ASGD's longer TTA, increasing 44-351\% and 38-427\% over SSGD.\looseness=-1




We group server resource records by the number of hosted PSs and plot the CDF of records versus average resource usage in Fig.~\ref{fig:cdf_res_ps_wrk_num}. As the number of PSs per server increases, both CPU and bandwidth usage rise, risking overload. When PS count grows from one to five, the share of CPU usage records above 98\% increases from 0\% to 2.3\%, and those above 90\% from 11\% to 100\%. Likewise, bandwidth records above 98\% rise from 0\% to 12\%, and above 90\% from 1\% to 65\%.\looseness=-1




\DEL{Fig.~\ref{subfig:cdf_mem_ps_num} and Fig.~\ref{subfig:cdf_mem_wrk_num} show 
that when the number of PSs hosted by the server changes from one to four, the percentage of records of memory usage that are greater than 15\% increases from 3\% to 12\%. When the number of workers hosted by the server changes from one to four to eight, the percentage of records of memory usage that are greater than 15\% increases from 4\% to 26\% to 90\%.
The reason why a server tends to have more memory usage with the increase in the number of workers than PSs is that a worker consumes higher memory usage than the PS for non-large models (\insref{ins_ps_worker_res}). Therefore, a server hosting more workers tends to be overloaded in CPU memory. However, since CPU memory is not a cause for stragglers.}





Fig.~\ref{fig:cdf_iter_ratio_for_ps_num} shows the CDF of iterations versus a worker's iteration time deviation ratio $d_i$ for four cases with different numbers of PSs on the worker's server. It suggests that an increase in the number of PSs on a worker's server correlates with an elevated ratio, indicating a higher likelihood of resource contention, longer iteration times, and straggler generation.\looseness=-1


\vspace{-0in}
\begin{ins}~\label{ins_ps_worker_res}
In both SSGD and ASGD, the PS consumes 5-87\% more CPU and 253-296\% more bandwidth than a worker. Servers hosting more PSs are prone to CPU and bandwidth overload, potentially causing stragglers among their hosted workers.
\end{ins}\vspace{-0in}

Since ASGD consumes significantly more bandwidth than SSGD, we examine whether switching to ASGD introduces additional stragglers. We co-located job A's (DenseNet121) PS with one worker each from jobs B and C (both MobileNet) and switched job A to ASGD during training. Fig.~\ref{subfig:switch_srv_res} shows that after switching, server CPU usage rose from 14-37\% to 46-58\%, and bandwidth from 50-90\% to 60-100\%. Fig.~\ref{subfig:col_wrk_iter_time} indicates job B's worker iteration time increased from 600-1200ms to 800-1600ms, and job C's from 200-1000ms to 690-1320ms. Fig.~\ref{subfig:switch_strg_num} further shows that while no stragglers existed before switching, both workers became frequent stragglers afterward.\looseness=-1



\vspace{-0in}
\begin{ins}~\label{asgd_straggler}
A job in ASGD uses 44\%-351\% more CPU and 38\%-427\% more bandwidth resources than SSGD. Therefore, when a job switches to ASGD, it can potentially create more stragglers in other jobs whose workers are co-located with the job's PS.
\end{ins}\vspace{-0in}

\DEL{In the single-job experiment, {\color{red}to compare the CPU consumptions for parameter update on the PS in SSGD and ASGD,} we chose the job ResNet56 and collected the CPU usage of the PS for each parameter update in SSGD and ASGD. {\color{red} For SSGD, we recorded the CPU usage of the PS for each parameter update (including gradient aggregation) every iteration from the start of gradient aggregation to the end of parameter update. For ASGD, we recorded the CPU usage of the PS for four parameter updates (i.e., the sum of four updates' CPU usage)
In the rest of the paper, we call one step one mini-batch unless otherwise specified.}
Fig.~\ref{fig:ps_cpu_ssgd_asgd} shows the CPU usage of the PS in SSGD and ASGD every four steps. We observe the CPU usage of the PS under ASGD is around 72\%, and the CPU usage of the PS under SSGD is around 20\%. {\color{red} The reason why CPU usage in SSGD is low is that the PS only updates parameters once every four steps (one iteration). Before each parameter update, there are three gradient addition operations under SSGD to aggregate the gradient copies from four workers. Under ASGD, the PS updates parameters four times every four steps. One parameter update takes time 29ms and consumes CPU usage 21\%. One gradient addition operation takes time 1.3ms and consumes CPU usage 16\%. So the parameter update dominates the CPU consumption in SSGD. {\color{red} The PS's CPU usage in SSGD is thus $(29ms\times 29\%+3\times 1.3ms\times 16\%)/(21ms + 3\times 1.3ms)\approx 20\%$.} Therefore, the CPU resources for four parameter updates are higher than for one parameter update and three gradient addition operations. The PS has higher CPU usage in ASGD than in SSGD.}


\begin{ins}~\label{ins_asgd_more_ps_cpu_comp}
The PS needs more CPU resources for computation in ASGD than in SSGD every $N$ steps, where $N$ is the number of workers. {\color{red}It indicates switching from SSGD to ASGD could overload the server in CPU, which is considered in Section~\ref{subsec:consider_asgd}.} ({\color{red} idea 4.c. Design 3.a uses it.}) (\DONE??show me where in my outline paper shows this, where in the design we used it)
\end{ins}

\begin{figure}[H]
\centering
\includegraphics[width=0.95\columnwidth]{figures/ps_cpu_ssgd_asgd_2.pdf}
    \caption{The PS's CPU usage for updating in SSGD and ASGD.}
\label{fig:ps_cpu_ssgd_asgd}
\end{figure}
}


\vspace{-0in}
\subsection{Effect of Switching to ASGD}\label{observe:acc_gain_stage}
\vspace{-0in}



Fig.~\ref{fig:jct_worker_cpu_strategy} shows the TTA of each model when we did not throttle, and throttled the CPU of worker1 to 75\%, 10\%, and 5\% of its capacity, respectively, under SSGD and ASGD. We see that a straggler barely affects TTA in ASGD but significantly increases TTA in SSGD, and higher throttling degree leads to higher TTA generally. When there is no straggler, SSGD has 9-127\% lower TTA than ASGD. When worker1's CPU is throttled to 75\% and 10\%, some jobs have 1\%-4\% and 6\%-18\% higher TTA in SSGD than in ASGD, while other jobs have 5\%-47\% and 4\%-36\% lower TTA in SSGD than in ASGD. When worker1's CPU is throttled to 5\%, all jobs have 3\%-61\% higher TTA in SSGD than in ASGD. This is caused by different model types and model sizes. Fig.~\ref{fig:jct_worker_bw_strategy1} illustrates similar observations when throttling worker1's bandwidth.




\begin{table}[h]\vspace{-0.35in}
\vspace{0.12in}
\caption{Accuracy improvement at different stage.} 
\vspace{-0.1in}
\label{tab:acc_gain_stage}
\centering
\footnotesize
\begin{tabular}{| c | c | c | c |}
    \hline
    & Step 2200 & Step 5500 &  Step 13000 \\
    \hline
    SSGDw/oS & 1.52\% & 0.64\% & 0.09\% \\
    \hline
    SSGDw/S & 0.48\% & 0.21\% & 0.02\% \\
    \hline
    ASGDw/S & 1.04\% & 0.29\% & 0.06\% \\
    \hline
\end{tabular}
\vspace{-0.15in}
\end{table}

Next, we check if switching to ASGD increases accuracy improvement, whether this increase is affected by training stages. We conducted two single-job SSGD experiments with and without creating a straggler by throttling worker1's CPU usage to 20\% of its capacity, denoted by SSGDw/S and SSGDw/oS.
Then, we conducted three SSGDw/S experiments but switched SSGD to ASGD (denoted by ASGDw/S) at step 2200 (early stage), step 5500 (middle stage), and step 13000 (late stage), respectively. We then measured the accuracy improvement gained in 2 minutes from the switching time. We selected 2 minutes to ensure observable accuracy improvements after switching while still reflecting the effect of the switching. The results for DenseNet121 are shown in Table~\ref{tab:acc_gain_stage} and similar observations can be made for other models. At the early, middle, and late stages, ASGDw/S exhibits 0.56\%, 0.08\%, and 0.04\% more accuracy improvement than SSGDw/S, respectively. SSGDw/S produces 1.04\%, 0.43\%, and 0.07\% lower accuracy improvement than SSGDw/oS.
As training progresses, stragglers lead to diminishing decrease in accuracy improvement in SSGD.\looseness=-1


\vspace{-0in}
\begin{ins}~\label{ins_tta} 
When no stragglers occur, SSGD has lower TTA than ASGD. When there are stragglers, switching to ASGD does not necessarily reduce TTA and its benefit varies among jobs and is influenced by model types, straggling degree and training stage.\looseness=-1

\end{ins}\vspace{-0in}

\begin{figure}
    \centering
    \subfigure[DenseNet121.\label{fig:sub:acc_diff_params_densenet121}]{\includegraphics[width=0.465\columnwidth,height=2cm]{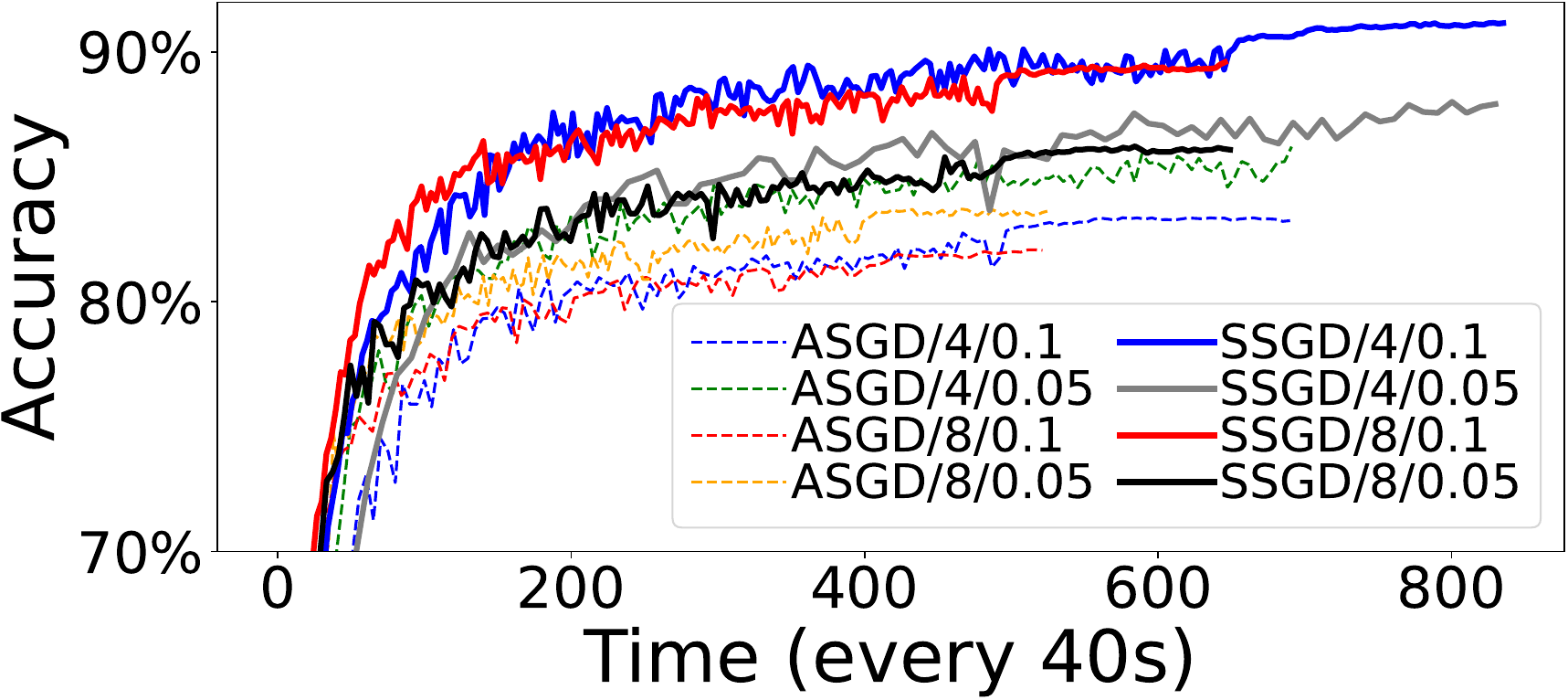}}
    \subfigure[LSTM\label{fig:sub:acc_diff_params_lstm}]{\includegraphics[width=0.465\columnwidth,height=2cm]{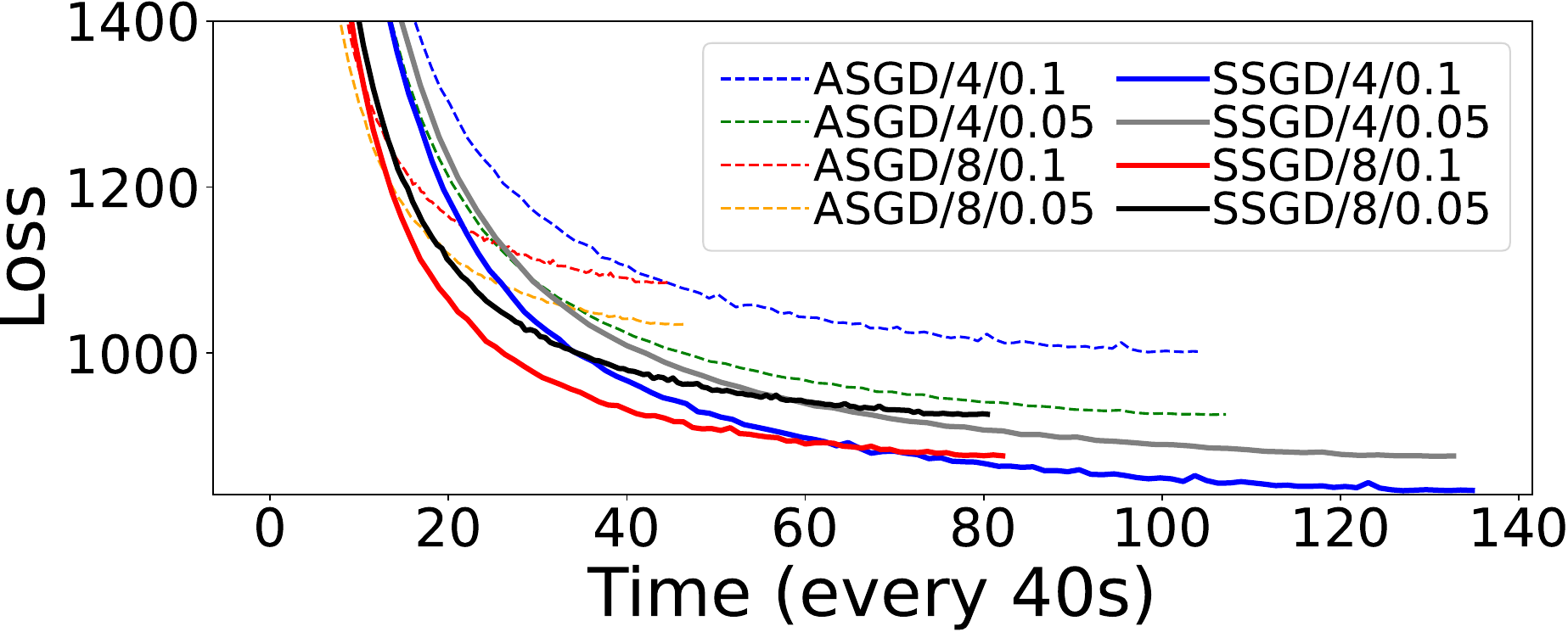}}
 \vspace{-0.1in}   \caption{Accuracy with different learning rates.}
    \label{fig:acc_diff_params1}\vspace{-0.2in}
\end{figure}

Next, we aim to investigate whether the optimal learning rate for SSGD is still optimal for ASGD after switching.
Fig.~\ref{fig:acc_diff_params1} shows the accuracy and perplexity over time of DenseNet121 and LSTM until they converge. We use /4 and /8 to represent 4 and 8 workers, and use /0.05 and /0.1 to represent a learning rate of 0.05 and 0.1. The results show that for SSGD, a learning rate of 0.1 achieves 2.8\%-3.1\% higher converged accuracy and 17-26 lower converged perplexity compared to 0.05. However, for ASGD, the optimal learning rate shifts, with 0.05 yielding 1.8\%-2.3\% higher accuracy and 96-107 lower perplexity than 0.1. The convergence results are influenced by the number of workers and demonstrate variations between SSGD and ASGD. Additional results for other models show consistent trends and are omitted due to space constraints.

\DEL{The figures for other models are in \cref{app:LR} which demonstrate similar results.
For the case of 4 workers, the achieved converged accuracy/perplexity and the training time for SSGD/0.1, SSGD/0.05, ASGD/0.1, and ASGD/0.05 are (90.7\%, 33520s), (87.9\%, 33360s), (83.6\%, 27600s), and (85.4\%, 27680s) for DenseNet, and (819, 5380s), (836, 5280s), (1104, 4300s), and (997, 4360s) for LSTM. For the case of 8 workers, they are (89.8\%, 25960s), (86.7\%, 26000s), (82.3\%, 20920s), and (84.6\%, 21040s) for DenseNet, and (832, 3380s), (858, 3320s), (1127, 1840s), and (1031, 1880s) for LSTM. For the models, SSGD/0.1 achieves 2.8\%-3.1\% higher converged accuracy and 17-26 lower converged perplexity than SSGD/0.05, but ASGD/0.05 achieves 1.8\%-2.3\% higher converged accuracy and 96-107 lower converged perplexity than ASGD/0.1. }

 \vspace{-0in}
\begin{ins}~\label{ins_optimal}
After switching to ASGD, the optimal learning rate of SSGD may not remain optimal for ASGD.
\end{ins}
 \vspace{-0in}


\vspace{-0in}
\section{System Design of STAR}\label{sec:design}
\vspace{-0in}

\insref{ins_straggler} emphasizes the importance of addressing stragglers in DL training on homogeneous GPUs. However, existing methods that switch to ASGD when a straggler is detected~\cite{li2021sync, damaskinos2018asynchronous, lian2018asynchronous} may generate more stragglers (\insref{asgd_straggler}), and its benefits depend on several factors (\insref{ins_tta}). In light of these challenges, we propose STAR, which introduces new synchronization modes and identifies the optimal synchronization mode to minimize TTA in the presence of a straggler. STAR incorporates the following based on insights gained from our observations. \looseness=-1


\squishlist
\vspace{-0.008in}\item[(1)] {Straggler prediction (\insref{ins_straggler}, \insref{ins_predict_ratio}) (\cref{subsec:straggler_detection}).}
\vspace{-0.008in}\item[(2)]{Static and dynamic $x$-order synchronization modes (\insref{ins_x_by_x}) (\cref{synch}).}\looseness=-1
\vspace{-0.008in}\item[(3)] {Synchronization mode determination (\insref{ins_tta}, \insref{ins_optimal}) (\cref{subsec:choose}).}
\begin{itemize}
\vspace{-0.008in}\item {Heuristic method (\cref{subsec:heuristic}).}
\vspace{-0.008in}\item {ML-based method (\cref{subsec:ml_mode}).}
\end{itemize}
\vspace{-0.008in}\item[(4)] {Resource-aware straggler prevention} (\cref{sec:resource}).
\begin{itemize}
\vspace{-0.008in}\item {Preventing stragglers upon mode change} (\insref{asgd_straggler}) (\cref{sec:resource1}).
\vspace{-0.008in}\item {Proactively preventing stragglers} (\insref{ins_straggler}, \insref{ins_ps_worker_res}) (\cref{sec:proactive}a).
\end{itemize}
\squishend

\begin{figure}
\includegraphics[width=0.985\columnwidth,height=2cm]{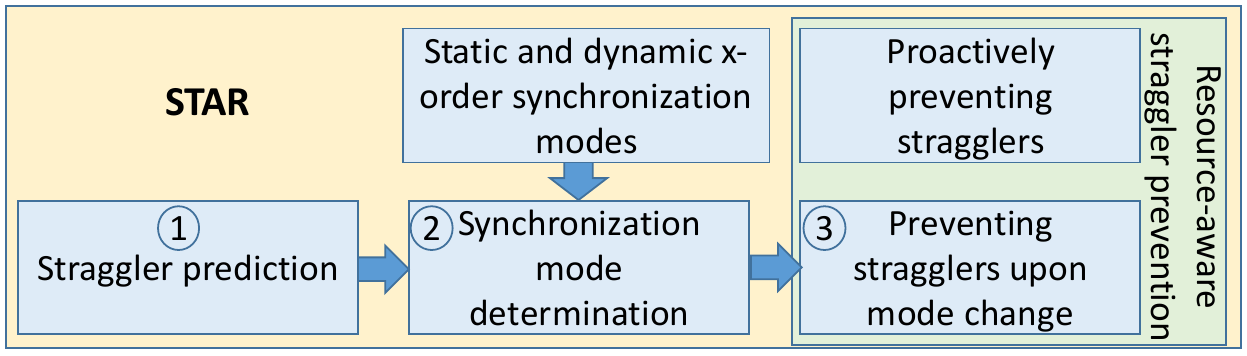}
\vspace{-0.15in}\caption{Overview of STAR.}
\label{fig:overview}\vspace{-0.2in}
\end{figure}

Fig.~\ref{fig:overview} shows an overview of the STAR system. At every iteration, STAR conducts straggler prediction ($\circled{1}$). If there are predicted stragglers, STAR determines the optimal synchronization mode for minimizing TTA ($\circled{2}$) and strategically reassigns resources between tasks to prevent stragglers caused by the mode change ($\circled{3}$) for the next iteration. Otherwise, the SSGD mode is used. In addition, STAR takes proactive measures to prevent stragglers by distributing the CPU and communication load among servers during task assignment and worker communication.

 \DEL{more accurately predicts stragglers and also judges TTA decrease from switching to ASGD or to ? to decide whether switch is needed and which synchronization mode is better. Specifically, we proposed a heuristic method and a ML method to determine synchronization mode when there are stragglers. If there will be stragglers in the next iteration, we use the heuristic/ML method to determine the mode. 
The heuristic method for mode determination takes a certain time for decision making while the ML method can quickly make a decision but needs a well trained model. Therefore, the system uses the heuristic method and uses the data for training the ML model, and then switches to use the ML method after the model is well trained. The model is kept being trained in the system. Based on \insref{ins_straggler_metric1} and \insref{ins_ps_worker_res}, STAR proactively schedules PSs to servers to avoid server overload or underload, and also forms workers and the PS of a job into a tree for communication to avoid stragglers caused by bandwidth.}

\vspace{-0.1in}
\subsection{Straggler Prediction}\label{subsec:straggler_detection}
\vspace{-0in}

\DEL{Based on \insref{ins_predict_ratio} and guided by~\insref{ins_straggler}, we use the available CPU and bandwidth of each worker to predict stragglers and iteration time. In addition, based on \insref{ins_straggler_Itimes}, we predict the time that there will be at least one straggler. 
So we do not need to check if these workers will be stragglers again in each iteration during this predicted time period.}



Motivated by the need of accurate straggler prediction (\insref{ins_predict_ratio}) and guided by the insights on straggler causes (\insref{ins_straggler}), we use the available CPU and bandwidth of each worker to predict iteration times and then stragglers. First, each worker predicts its received CPU and bandwidth in the next iteration using LSTM based on the historical data from the last $n$ (e.g., 100) iterations. Utilizing the predicted received CPU/bandwidth, along with information on model type and batch size as inputs, each worker employs a regression model to predict its iteration time and computation completion time, and sends the information to the PS or a selected proxy in the AR architecture by piggybacking it to the gradients. The PS/proxy calculates the iteration time deviation ratio, identifies potential stragglers, and determines the optimal synchronization mode if stragglers are detected (as described below).\looseness=-1


\vspace{-0.1in}
 \subsection{Static and Dynamic x-order Synchronization Modes} \label{synch}
 \vspace{-0in}



Considering the trade-offs between ASGD and SSGD (explained in \cref{sec:Intro}), based on the rationale in \insref{ins_x_by_x}, we propose a static $x$-order synchronization mode, where parameters are updated using the gradients from $x$ ($1$$<$$x$$<$$N$) workers each time. A higher order synchronization mode achieves a higher converged accuracy and lower TTA since more gradients being synchronized reduces stale gradients but may increase TTA with stragglers.
To verify this, in a single-job experiment with 8 workers, we applied 1-order, 2-order, 4-order, and 8-order synchronization modes to all models until convergence. The results, illustrated in Fig.~\ref{fig:asgd_x_by_x_time}, indicate that 1-order, 2-order, 4-order, and 8-order synchronization modes achieve converged accuracies of 80.3\%, 82.7\%, 86.4\%, and 88.9\%, with corresponding TTA values of 15680s, 4120s, 2480s, and 1960s, respectively.

\begin{wrapfigure}{c}{4.1cm}\vspace{-0.1in}
\includegraphics[width=0.46\columnwidth,height=2.3cm]{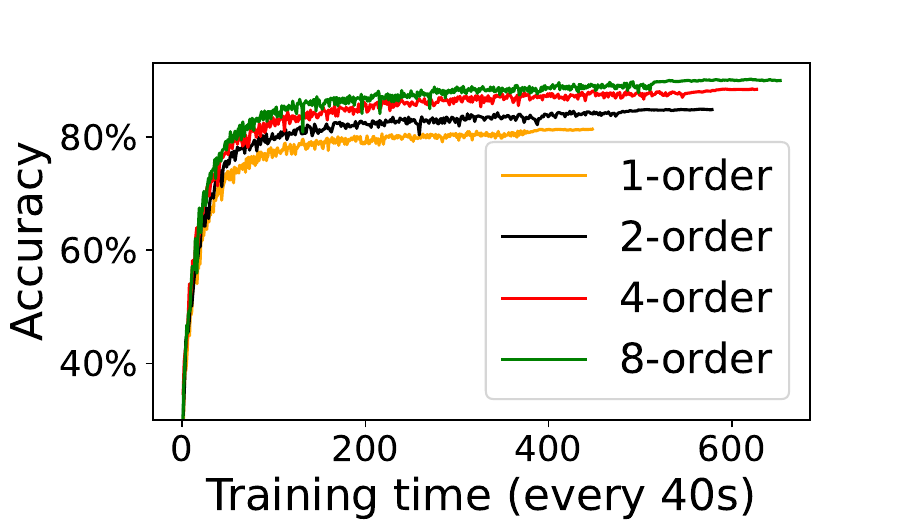}
\vspace{-0.3in}\caption{Accuracy.}
\label{fig:asgd_x_by_x_time}\vspace{-0.23in}
\end{wrapfigure}

Based on \insref{ins_x_by_x}, we further propose a dynamic-$x$-order synchronization mode that clusters workers with similar predicted iteration times using a clustering algorithm (e.g., agglomerative hierarchical clustering~\cite{agglomerative-clustering}). The PS then uses the gradient reports from each worker cluster to update parameters. This mode aims to enhance the efficiency of parameter updates by adapting to the varying conditions of worker iteration times, thereby reducing the PS waiting time compared to the static-$x$-order synchronization mode.


%

\noindent \textbf{All-reduce architecture.}  Suppose there are $X$ stragglers, and our strategy is to remove $x$ stragglers (where $x\leq X$) and connect each of them with a non-straggler (as its parent) that has high bandwidth to it to report its gradients and receive parameters. A parent waits for a certain time ($t_w$) after its computation before aggregating the gradients from its children and its own gradients, and then broadcasts the aggregated gradients. Different values of $x$ and $t_w$ form different synchronization modes.

\vspace{-0.05in}
\subsection{Synchronization Mode Determination}\label{subsec:choose}
\vspace{-0in}
\subsubsection{Heuristic Method}\label{subsec:heuristic}\vspace{-0in}






The heuristic method selects the optimal synchronization mode among various options, including SSGD, ASGD, static $x$-order modes ($x=2, ..., N-1$) and dynamic $x$-order modes). The goal is to minimize the time required to achieve a certain \emph{training progress}, which can be reflected by the accuracy improvement. To calculate this time for the static-$x$-order modes (denoted by $T_x$), we first determine the number of parameter updates needed for each synchronization mode to achieve the same training progress (denoted by $n_u$), and then multiply it with the expected time needed for one parameter update (denoted by $t_x$). $n_u$ considers the model type and training phase and $t_x$ considers the straggling degree (\insref{ins_tta}) in the synchronization mode selection. $n_u$ can be expressed by \textit{pre-conditioned gradient noise scale} (PGNS) \cite{aurick2021pollux, sam2018empiricalmodel}. PGNS increases as the model becomes more accurate during training, indicating that more parameter updates are required to achieve a certain progress \cite{sam2018empiricalmodel}. The PGNS for the $k$-th parameter update is expressed as $\varphi_k = \frac{\mbox{tr}(P\sum P^T)}{|Pg|^2}$, where $tr()$ is the trace of a matrix defined as the sum of its diagonal elements, $P$ is the pre-conditioning matrix used by SGD with a loss function of $\mathcal L(Pw)$ (where $w$ represents weights), $\sum$ is the covariance matrix of the gradients over each single data sample~\cite{sam2018empiricalmodel}, and $g$ is the true gradient over all data samples~\cite{aurick2021pollux}. $n_u$ for total batch size $M$ summed across all workers is equal to $(1 + \frac{\varphi_k}{xM/N})$~\cite{sam2018empiricalmodel}. Based on the predicted iteration times of all workers (from \cref{subsec:straggler_detection}), we calculate $t_x$. Finally, $T_x$ is computed as follows: \looseness=-1\vspace{-0.05in}
\begin{equation}
        T_x = (1 + \frac{\varphi_k}{xM/N})\times t_x. \vspace{-0.05in}
\end{equation}

In the dynamic-$x$-order mode, we first sort the clusters in the ascending order of the maximum iteration time $t_{c_i}$ in each cluster, denoted by $\{c_1, c_2, ..., c_{N_c}\}$, where $N_c$ is the number of clusters. Cluster $c_i$ has $n_{c_i}$ workers. Then, we calculate the training progress achieved by $c_i$ per unit time: $\frac{1}{(1+\frac{\varphi_k}{n_{c_i}M/N})\times t_{c_i}}$. Therefore, the expected time to achieve the same training progress in the dynamic-$x$-order synchronization mode equals: \vspace{-0.15in}
\begin{equation}\label{equ:progress_time_group}
        T_d = \frac{1}{\sum_{i=1}^{N_c}\frac{1}{(1+\frac{\varphi_k}{n_{c_i}M/N})\times t_{c_i}}}, \vspace{-0.05in}
\end{equation}Finally, we choose the synchronization mode that has the least time among $T_x$ ($x=1, 2, ..., N$) and $T_d$ to minimize TTA.\looseness=-1

Calculating PGNS in real time from scratch is infeasible because the covariance matrix of per-sample gradients $\sum$ takes hours to calculate particularly for large datasets. To address this, \cite{aurick2021pollux} approximates $\varphi_k$ by using a model's pre-calculated PGNS for the $e^{th}$ epoch ($\varphi_e$). To make this approach applicable to our synchronization modes, we extend this approach by pre-calculating $\varphi_s$ at intervals of $s$ steps of an iteration. The current PGNS is then approximated as $\varphi_s$, where $s$ is the nearest completed number of steps.

\noindent \textbf{All-reduce architecture.}
To select the optimal synchronization mode for the AR architecture, we need to find the values of the number of removed workers ($x$) and parent wait time ($t_w$) that minimize the time required to achieve a certain training progress. We calculate the time by: \vspace{-0.15in}

\vspace{-0.1in}
\begin{equation}\label{eq:all_reduce_strg_remove}
        T_a = (1+\frac{\varphi_k}{(N-x+q)M/N})\times (t_{ring}+t_w), \vspace{-0.06in}
\end{equation}where $t_{ring}$ is the maximum iteration time among the workers in the ring, and $q$ is the number of removed stragglers that have new iteration times no higher than $t_w$. We enumerate the values of $x$ and $t_w$ in their reasonable ranges and find their values that minimize $T_a$.

\noindent\textbf{Scaling learning rate after switching.} When switching from SSGD to another synchronization mode, the batch size for one parameter update is reduced from $M$ to $M_{new}=yM/N$, where $y$ is the number of gradient reports used for the update. Motivated by \insref{ins_optimal} and based on~\cite{priya2017accurate,christopher2018measuring}, we scale the optimal learning rate of SSGD, denoted as $r_{SSGD}$, proportionally with the batch size by setting $r_{new} = \frac{M_{new}}{M}r_{SSGD}$.\looseness=-1  

\vspace{-0in}
\subsubsection{ML-based Method}\label{subsec:ml_mode}
\vspace{-0in}


The heuristic method requires a certain amount of time (e.g., $\sim$970ms in \cref{sec:exp}) to make a decision and lacks pre-calculated PGNSs for new models. To address the issues, we propose a regression-based method. STAR initially employs the heuristic, gathers data to train the regressor, and subsequently transitions to using the trained regressor while continuing to refine the model through training.\looseness=-1



The regressor predicts the time latencies to achieve the same training progress for different synchronization modes. STAR then selects the mode with the shortest time latency. The inputs to the regressor include the predicted iteration time of each worker, the deviation ratio $\frac{T_{i}-\min_{i=1}^{N} T_{i}}{\min_{i=1}^{N} T_{i}}$, the model type, the learning rate, and the training stage (represented by the number of completed steps) based on insights from \insref{ins_tta} and \insref{ins_optimal}. Additional results (not shown due to space limitations) confirm the importance of the inputs, as indicated by their significant scores. 





\vspace{-0.05in}
\subsection{Resource-aware Straggler Prevention}\label{sec:resource}

\subsubsection{Preventing Stragglers Upon Mode Change}\label{sec:resource1}

We use the term ``task'' to denote both PSs and workers. For a job's selected synchronization mode, predicted stragglers may cause co-located jobs to slow down (\insref{asgd_straggler}). In AR, parents with higher bandwidth and CPU usage can similarly induce stragglers in co-located jobs. Thus, we verify whether each task has sufficient CPU and bandwidth. If not, we re-assign resources from other co-located tasks to meet its needs while minimizing their impact. For an $x$-worker group, if the slowest worker (in PS) or parent (in AR) completes at time $t$, faster peers need not finish earlier since this does not affect TTA. Likewise, in AR, if the slowest child's gradients reach its parent at time $t$, other stragglers need not finish beforehand. Hence, we re-assign CPU and bandwidth from these co-located workers, delaying their execution so they complete at $t$.\looseness=-1


If this approach still fails to satisfy the task's resource demands, we distribute the inadequate amount of type-$k$ resource (denoted by $R^k$) among co-located tasks based on their sensitivity to the type-$k$ resources and training phase, represented by the current accuracy improvement (\insref{ins_tta}). A co-located task with a lower sensitivity to type-$k$ resource (denoted by $S^k$) and lower current accuracy improvement (denoted by $A$) can afford more resource reduction, and vice versa. The sensitivity is $\Pi\frac{TTA_j^k-TAA}{TTA}$, where $TTA_j^k$ is the TTA in the $j^{th}$ throttling of the type-$k$ resource and $TTA$ is that with no resource throttling. The amount of $k$-type resource reduced from a task $i$ equals $R^k \times \frac{1/(S_i^k\cdot A_i)}{\sum_i 1/(S_i^k\cdot A_i)}$. 
Recall we have a regression model that predicts iteration time and computation completion time based on available CPU/bandwidth.
By using it, we calculate the sum of iteration times of all impacted jobs and this job with and without the resource reassignment, denoted by $S_w$ and $S_o$, respectively. If $S_w<S_o$, we use the resource reassignment. Otherwise, we choose the next synchronization mode with the second-lowest time to achieve the same training progress and repeat this process until we find a solution with enough resources to execute.\looseness=-1


Based on \insref{ins_ps_worker_res}, a worker's CPU and bandwidth consumption is much less than a PS's and it may not overload its server. In the PS architecture, STAR thus does not have to consider the resource demands of the job's workers here to save overhead (\cref{sec:exp} shows the results with and without this consideration).\looseness=-1

\vspace{-0in}
\subsubsection{Proactively Preventing Stragglers}\label{sec:proactive}

Beyond preventing stragglers from synchronization mode changes, we proactively avoid those from server load imbalance caused by high-load tasks such as PSs and parents in AR. Servers hosting multiple PSs often trigger stragglers (\insref{ins_ps_worker_res}), while both overloaded and underutilized servers can also create them (\insref{ins_straggler}). In AR, parents with many children impose heavier CPU and bandwidth demands, which may induce stragglers on their servers.\looseness=-1

\paragraph{High-load Task Assignment}\label{sec:guid:job_cpu_gpu_padding}\vspace{-0in}


We ensure PSs are evenly distributed across servers. Although Muri~\cite{zhao2022interleaving} can assist, it assumes periodic resource usage, which may not hold under stragglers. Thus, we design a simple heuristic that balances the number of PSs per server. When assigning a PS using prior scheduling methods (e.g.,~\cite{zhao2022interleaving}), we prioritize servers capable of hosting more PSs given available CPU and bandwidth. In AR, we likewise balance child assignments by letting each child select a high-bandwidth ring worker with the fewest existing children. By preventing stragglers, STAR helps restore the periodicity assumption and preserves the effectiveness of prior scheduling methods.\looseness=-1

\DEL{Though the resource utilizations of
PSs exhibit periodical patterns, because of the straggling, the
resource usage maybe shiftedlater,whichmaycause resource
contention with other PSs.

For this purpose, we leverage the feature that the resource consumption of a PS follows certain pattern (??add Yibo zhu's sigcomm paper reerence) (\insref{ins_res_pattern} from Appendix).


\DEL{ML training consumes a large amount of resources in the system~\cite{abadi2016tensorflow}. Resource contention can cause stragglers. In the system level, if we fully utilize resources, we can avoid stragglers, reduce TTA, and enable the system to run more jobs on limited resources. In this paper, we leverage the unique patterns of resource consumption of a PS and a worker to fully utilize resources and improve TTA.}


Given a group of jobs to be scheduled, we first estimate the CPU active and idle time of the PS. 
For known jobs, we directly use the historical data. For unknown jobs we perform a few dry runs to generate the data. 
We aim to allocate some PSs together in a server so the CPU and bandwidth are not over-utilized or under-utilized (since both could be the causes for stragglers). Though the resource utilizations of PSs exhibit periodical patterns, because of the straggling, the resource usage may be shifted later, which may cause resource contention with other PSs. Thus, we assign $D$ more time to a PS, where $D$ is the average delay caused by straggling in the historical data. A PS should be assigned to a server closer to its workers to reduce communication latency, and thus a few server options are determined. This is {\color{red} an NP-hard} bin-packing problem~\cite{lee1985bin-packing}. (\DONE?? what problem NP-hard or polynomial we need paper for complexity) ({\color{red} In muri paper}??show me Yibo's paper) Firstly grouping PSs and then assigning each group~\cite{zhao2022interleaving} also would take a long time {\color{red} $O(n^3log_2k)$, where $k$ is the number of PSs in each group}. We use a lightweight heuristic approach here. To choose a server to assign the PS, we choose the one that will not be overloaded by assigning the PS and also lead to the minimum resource under-utilization.


%


Given the PS of a job, in order to find a close-to-perfect matching, we use CPU-BW interleaving efficiency $\gamma$ of a job-server matching pair defined in~\cite{zhao2022interleaving} (i.e., the fraction of time that CPU and BW are not idle), which is expressed as
\begin{equation}
    \begin{split}
        \gamma = 1-\frac{1}{2}(\frac{T-t_0^{CPU}-t_1^{CPU}}{T}+\frac{T-t_0^{BW}-t_1^{BW}}{T}),
    \end{split}
\end{equation}

\noindent where $t_0^{CPU}$ and $t_0^{BW}$ are the CPU and BW active time of the job's PS, $t_1^{CPU}$ and $t_1^{BW}$ are the CPU and BW active time of the existing PSs of the server, and $T$ is the time of one iteration defined as

\begin{equation}
    \begin{split}
        T = max(t_0^{CPU},t_1^{BW}) + max(t_0^{BW},t_1^{CPU}).
    \end{split}
\end{equation}

Of course, we can also use the same approach to assign workers.
}

\DEL{\subsubsection{if not switch to ASGD, can jump in other PSs to use the resource (deadline) (??need to add)}\label{JumpIn}
\insref{ins_straggler_metric1}, \insref{ins_straggler_metric}, \insref{ins_ps_worker_res},

\insref{ins_ssgd_long_tta_low_res} to fully utilize resources

If our approach decides that there is no need to switch to ASGD, then the PS will wait. The CPU and bandwidth resources during this waiting time will not be used. In order to fully utilize these resources, we propose running a delay-tolerant task that uses CPU and bandwidth during this waiting time. Some tasks may be delay-tolerant such as data pre-processing or post-processing or some jobs have very low demand on TTA. These tasks can be classified as delay-tolerant tasks and only use such ideal resources when available.}




\paragraph{Amortizing Communication Overhead}\label{sec:tree} \vspace{-0in}

\insref{ins_straggler} shows varying communication latencies among workers. To mitigate stragglers from worker-PS or worker-parent communication, we distribute communication overhead via a tree structure with the PS or parent as the root. Workers with longer latencies are placed in lower layers and connected to higher-layer workers with low-latency links. Each worker aggregates its gradients with those from its children and forwards the result upward, overlapping communication and computation in a bottom-up fashion. The root updates parameters and sends them top-down. The PS (or a proxy in AR) constructs the tree before each iteration and informs workers of their upper-layer connections by piggybacking this information on parameters.\looseness=-1


\vspace{-0.05in}
\section{Performance Evaluation}\label{sec:exp}\vspace{-0.in}
\DEL{In this section we evaluated our proposed methods with regard to JCT, TTA, converged accuracy and investigated their time overhead.}

\subsection{Implementation and Setup}\vspace{-0in}
\noindent\textbf{Implementation.} We implemented STAR in PyTorch~\cite{paszke2019pytorch}, using its RPC for communication. Server and job resource usage were monitored via the Python \textit{psutil} package~\cite{psutil}, running in separate processes to avoid overhead. Data was stored in shared memory using Python's \textit{multiprocessing.shared\_memory}. Each job had a manager process to read resource data and manage threads for STAR model training, which ran on separate GPUs to avoid interfering with DL jobs. CPU and bandwidth control were enforced with \textit{cpulimit}~\cite{cpulimit-linux} and \textit{tc}~\cite{tc-linux}.

Experimental settings followed~\cref{sec:exp_measure} unless noted otherwise. Job completion time (JCT) denotes the convergence time of a job~\cite{li2021sync}.\looseness=-1

%

%

\noindent\textbf{Comparison methods.} We compared STAR with six existing systems: SSGD, ASGD, Zeno++~\cite{zeno++}, Live Gradient Compensation (LGC) \cite{xu2021lgc}, Sync-Switch \cite{li2021sync}, and LB-BSP \cite{chen2020lbbsp}. Zeno++ and LGC represent two typical variants of vanilla ASGD. Zeno++ is ASGD with bounded staleness, utilizing a small separate validation set to measure accuracy, and applies gradients for parameter updates only when the accuracy does not decrease.
In LGC, the gradients from the $K$ fastest workers (we set $K=5$) are used for the parameter update. Sync-Switch is a typical method that alternates between SSGD and ASGD. It identifies a worker as a straggler if straggling is observed for 5 seconds, switches to ASGD in the presence of predicted stragglers, and reverts to SSGD otherwise. LB-BSP is another typical method that mitigates stragglers by adjusting workload. If the fastest worker consistently achieves lower iteration time than the slowest worker for a certain number of iterations (we used 8 iterations), it adjusts their batch sizes by a certain amount (set to 32 samples).\looseness=-1

We adapted LGC for AR by excluding the $N-K$ slowest workers from the ring and connecting each to a ring worker with the highest bandwidth to it. We used Sync-Switch's source code~\cite{sync-swtich-code} and developed LB-BSP, LGC, and Zeno++ (due to unavailability of their source code). In AR, $x$ was varied from 1 to the number of stragglers, and $t_w$ was set to 30-210ms for different models. These parameter values were chosen empirically to achieve optimal performance.\looseness=-1

We trained the ML model for STAR-ML, which took around 1.7 hours, using data from several dry runs using STAR-H. To see the individual performance of STAR-H and STAR-ML, we used STAR-ML from the beginning of the experiment. One iteration takes 100-800ms for different models, while the heuristic decision-making takes around 970ms, which would cause a training pause until the
decision is made. So we also conducted decision-making 970ms before each iteration to avoid the pause, denoted
as STAR-.\looseness=-1 


\vspace{-0.05in}
\subsection{Overall Performance Comparison}\vspace{-0in}



\begin{wrapfigure}{c}{4.3cm}\vspace{-0.1in}
\centering
\includegraphics[width=0.5\columnwidth,height=2.5cm]{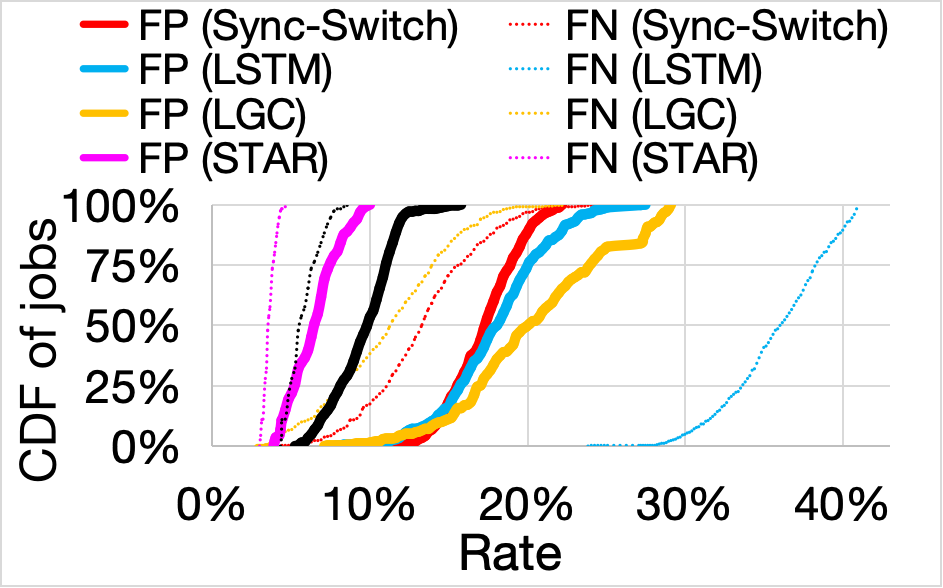}
\vspace{-0.3in}\caption{Straggler prediction accuracy.}\vspace{-0.2in}\label{fig:relwrk_false_posi_nega}
\end{wrapfigure}

\noindent\textbf{Straggler prediction.} Fig.~\ref{fig:relwrk_false_posi_nega} shows the CDF of jobs versus false positives (FPs) and false negatives (FNs) among all 350 jobs. STAR achieves the lowest FPs (3.5-10.4\%) and FNs (3.8-4.2\%) compared with other methods. STAR- predicts earlier in the iteration window, leading to higher FPs and FNs~\cite{gao2020machine}. These results indicate that identifying stragglers solely by fixed delays or slowest iteration times is imprecise.\looseness=-1

\vspace{-1pt}

\begin{figure}[h]\vspace{-0.1in}
\centering
\subfigure[PS architecture.]
{\includegraphics[width=0.49\columnwidth,height=2cm]{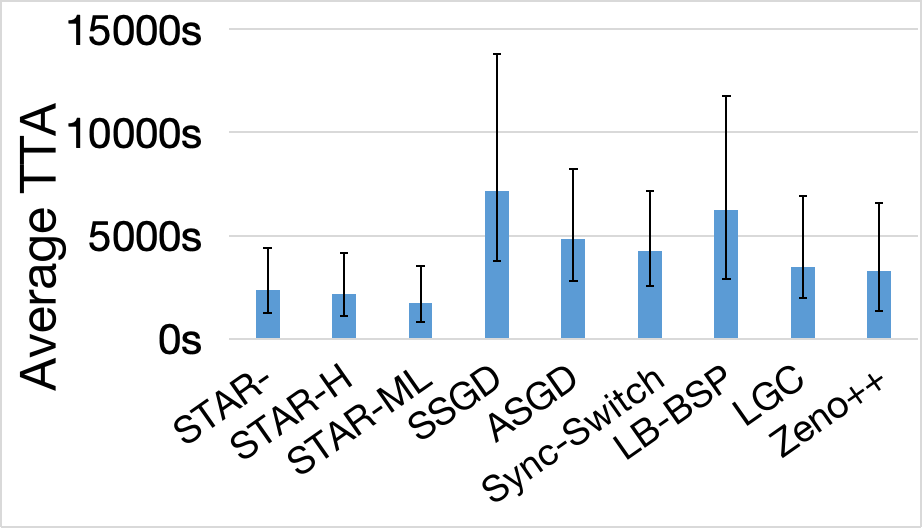}\label{subfig:concurr_avgtta_aws}}
\subfigure[All-reduce architecture.]
{\includegraphics[width=0.49\columnwidth,height=2cm]{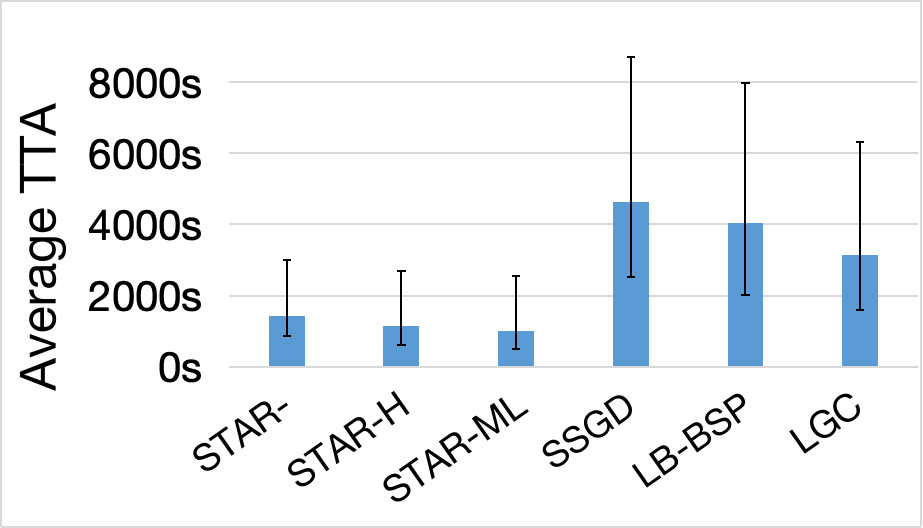}\label{subfig:concurr_avgtta_aws_allred}}
\vspace{-0.2in}\caption{TTA of each job in different systems.}\vspace{-0.1in}
\label{fig:concurr_avgtta}
\end{figure}

\noindent\textbf{TTA.} Fig.~\ref{fig:concurr_avgtta} shows the average TTA per job and the 1st and 99th percentiles across systems. In PS, STAR-ML achieves 84\%, 69\%, 62\%, 78\%, 52\%, and 48\% lower average TTA than SSGD, ASGD, Sync-Switch, LB-BSP, LGC, and Zeno++, respectively. STAR-H also reduces TTA by 77\%, 58\%, 51\%, 70\%, 42\%, and 36\%, benefiting from accurate straggler prediction, added $x$-order modes, optimal mode selection, and proactive avoidance. SSGD has the highest TTA due to straggler delays, while ASGD improves by skipping them. LB-BSP performs slightly better than SSGD by resizing mini-batches, and LGC, Zeno++, and Sync-Switch further reduce TTA via synchronous updates that avoid stale gradients.

STAR-ML's TTA is 18\% lower than STAR-H due to faster decision making, as its ML inference overlaps with training. In contrast, STAR-H's heuristic may exceed an iteration, briefly pausing training. Still, STAR-H's TTA is 7\% lower than STAR- since its improved decisions outweigh the pause overhead. Similar trends appear in Fig.~\ref{subfig:concurr_avgtta_aws_allred}. In AR, STAR-H achieves 66\%, 55\%, and 43\% lower TTA than SSGD, LB-BSP, and LGC, while STAR-ML reduces them by 70\%, 59\%, and 51\%. Systems with higher average TTAs also exhibit larger variances.\looseness=-1






\DEL{In Fig.~\ref{subfig:concurr_avgtta_aws_allred}, the relationship between different systems is the same as in Fig.~\ref{subfig:concurr_avgtta_aws} due to the same reasons. Compared to SSGD, LB-BSP, and LGC, STAR-ML produces 59\%, 53\%, and 39\% lower average TTA, STAR-H produces 57\%, 51\%, and 37\% lower average TTA, and STAR- produces 15\% higher average TTA than STAR-H. Both figures demonstrate that systems with higher average TTAs also have higher variances. \looseness=-1
}



\begin{figure}[h]\vspace{-0.1in}
\centering
\subfigure[PS architecture.]
{\includegraphics[width=0.49\columnwidth,height=2cm]{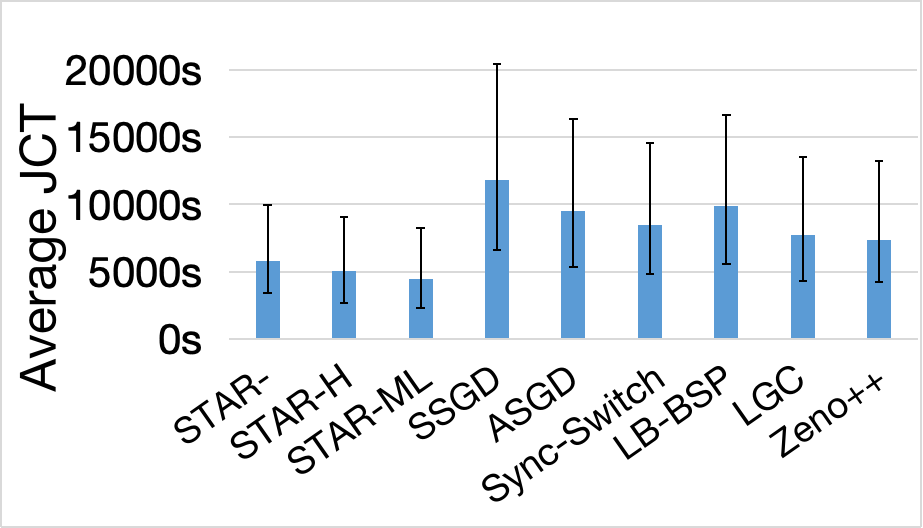}\label{subfig:concurr_avgjct_aws}}
\subfigure[All-reduce architecture.]
{\includegraphics[width=0.49\columnwidth,height=2cm]{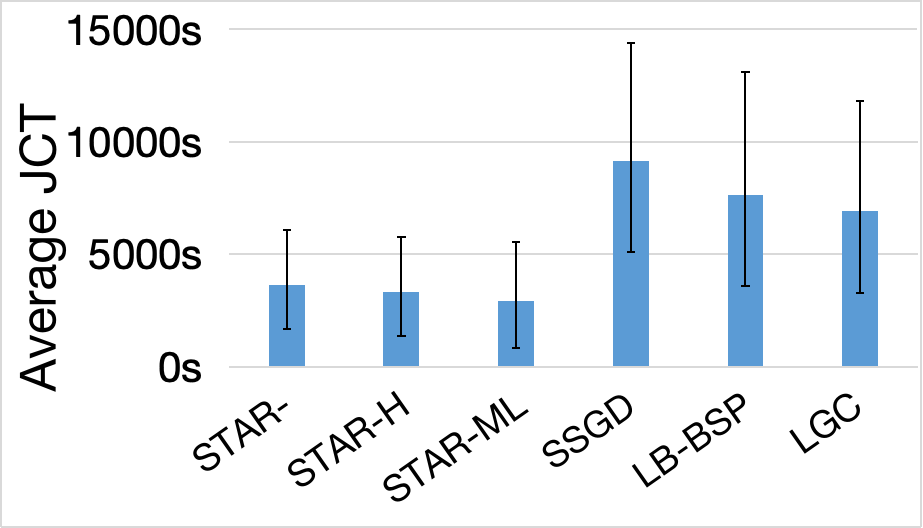}\label{subfig:concurr_avgjct_aws_allred}}
\vspace{-0.2in}\caption{JCT of each job in different systems.}\vspace{-0in}
\label{fig:concurr_avgjct}
\end{figure}

\noindent\textbf{JCT.} Fig.~\ref{fig:concurr_avgjct} shows the average JCT per job and the 1st and 99th percentiles across systems. In PS, STAR-H achieves 58\%, 45\%, 36\%, 47\%, 28\%, and 22\% lower average JCT than SSGD, ASGD, Sync-Switch, LB-BSP, LGC, and Zeno++, respectively, while STAR-ML further reduces them by 64\%, 48\%, 41\%, 55\%, 37\%, and 33\%. STAR-ML and STAR- yield 13\% lower and 16\% higher JCT than STAR-H, respectively. In AR, STAR-ML achieves 77\%, 64\%, and 55\% lower average JCT than SSGD, LB-BSP, and LGC, and STAR-H achieves 70\%, 61\%, and 52\% lower values. Both architectures show consistent trends with Fig.~\ref{fig:concurr_avgtta}, for the same underlying reasons.\looseness=-1

\begin{figure}[h]\vspace{-0.2in}
\centering
\subfigure[PS architecture.]
{\includegraphics[width=0.49\columnwidth,height=2cm]{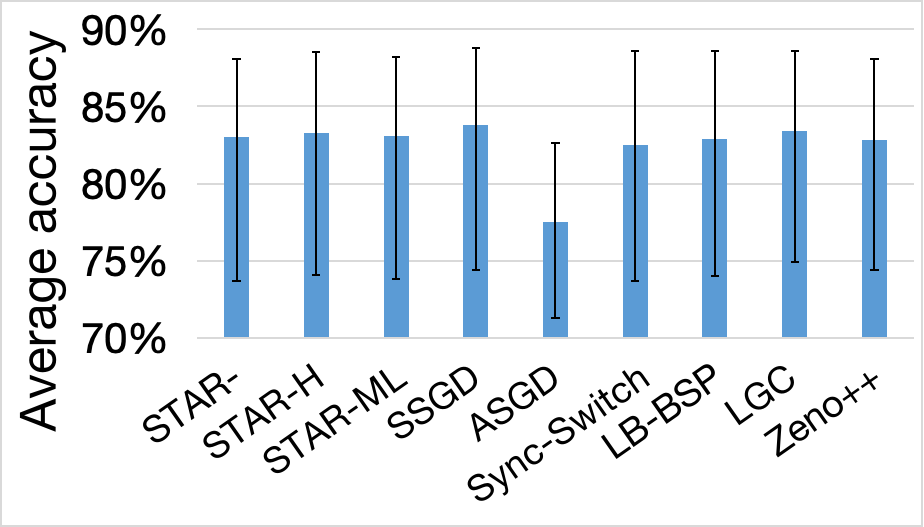}\label{subfig:concurr_avgacc_aws}}
\subfigure[All-reduce architecture.]
{\includegraphics[width=0.49\columnwidth,height=2cm]{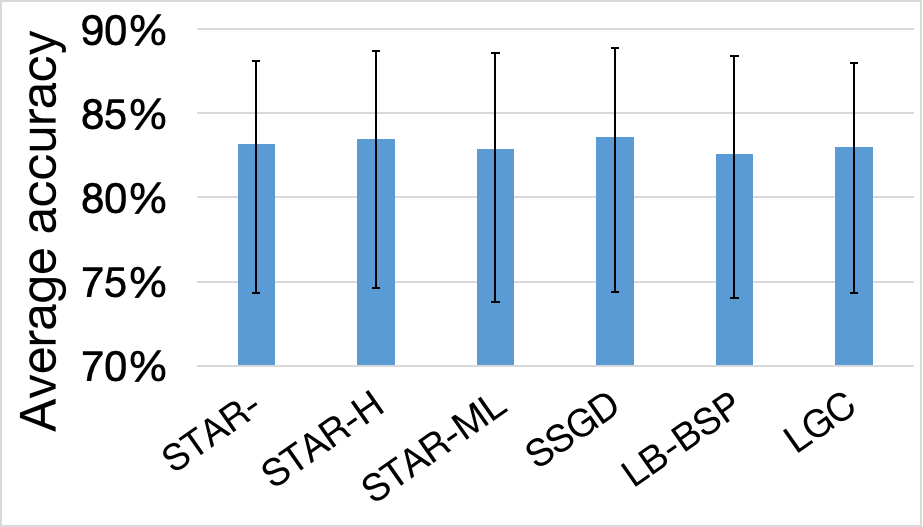}\label{subfig:concurr_avgacc_aws_allred}}
\vspace{-0.2in}\caption{Accuracy of image-classification jobs.}\vspace{0in}
\label{fig:concurr_avgacc}
\end{figure}

\begin{figure}[h]\vspace{-0.2in}
\centering
\subfigure[PS architecture.]
{\includegraphics[width=0.49\columnwidth,height=2cm]{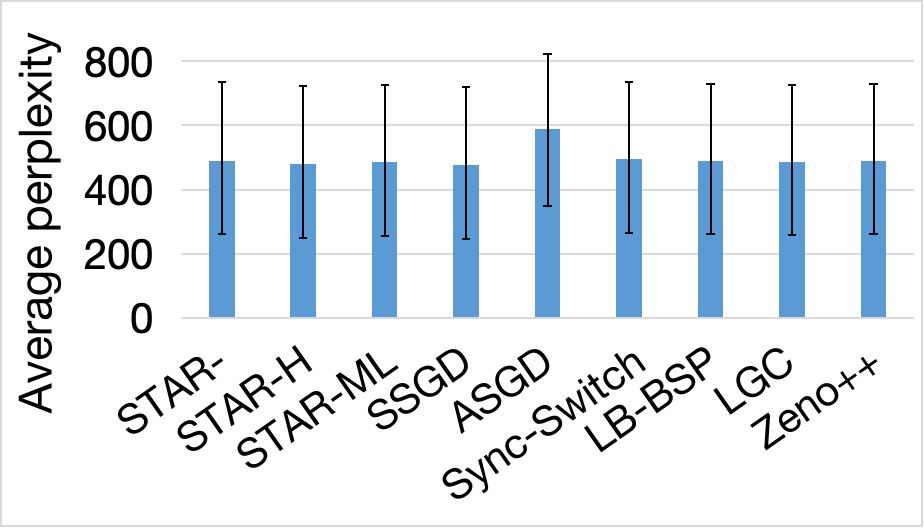}\label{subfig:concurr_avgperp_aws_lstm}}
\subfigure[All-reduce architecture.]
{\includegraphics[width=0.49\columnwidth,height=2cm]{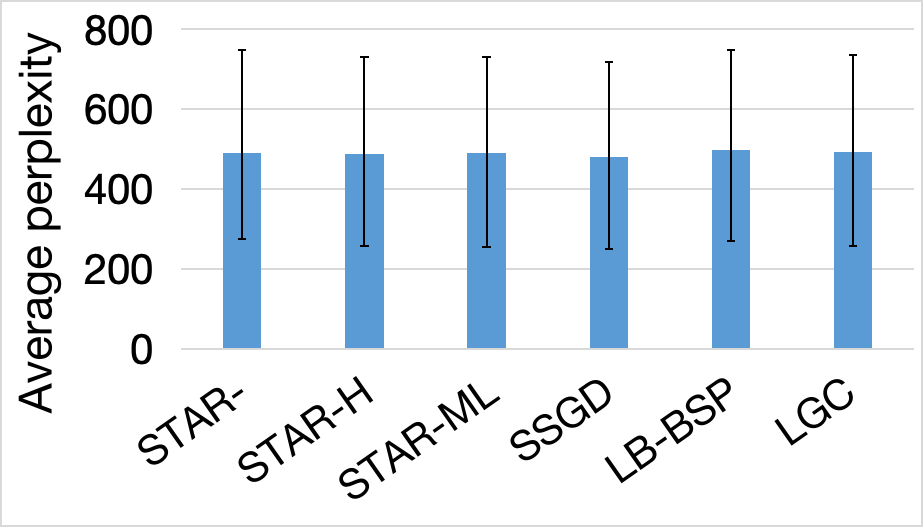}\label{subfig:concurr_avgperp_aws_transf}}
\vspace{-0.2in}\caption{Average perplexity of NLP jobs.}\vspace{-0.05in}
\label{fig:concurr_avgperp}
\end{figure}


\noindent\textbf{Converged accuracy and perplexity.} Fig.~\ref{fig:concurr_avgacc} shows the average converged accuracy per job and the 1st and 99th percentiles for image-classification tasks. In PS, STAR-H and STAR-ML achieve accuracy similar to SSGD (84\%) and about 1\% higher than other systems. The same trend appears in AR. Fig.~\ref{fig:concurr_avgperp} presents the average converged perplexity per NLP job, with relationships consistent with Fig.~\ref{fig:concurr_avgacc}. In all figures, higher average metrics correlate with greater variance.

\begin{figure}[h]\vspace{-0.1in}
\centering
\subfigure[PS architecture.]
{\includegraphics[width=0.49\columnwidth,height=2cm]{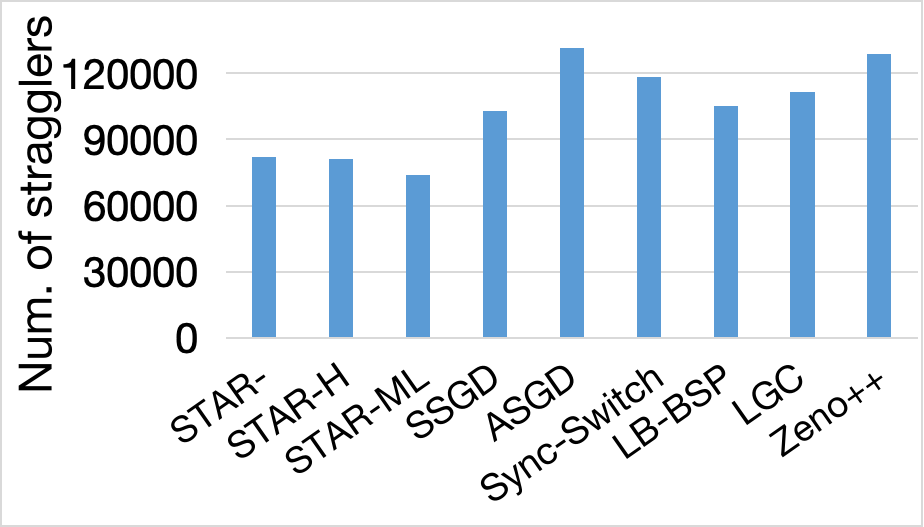}\label{subfig:concurr_strg_num}}
\subfigure[All-reduce architecture.]
{\includegraphics[width=0.49\columnwidth,height=2cm]{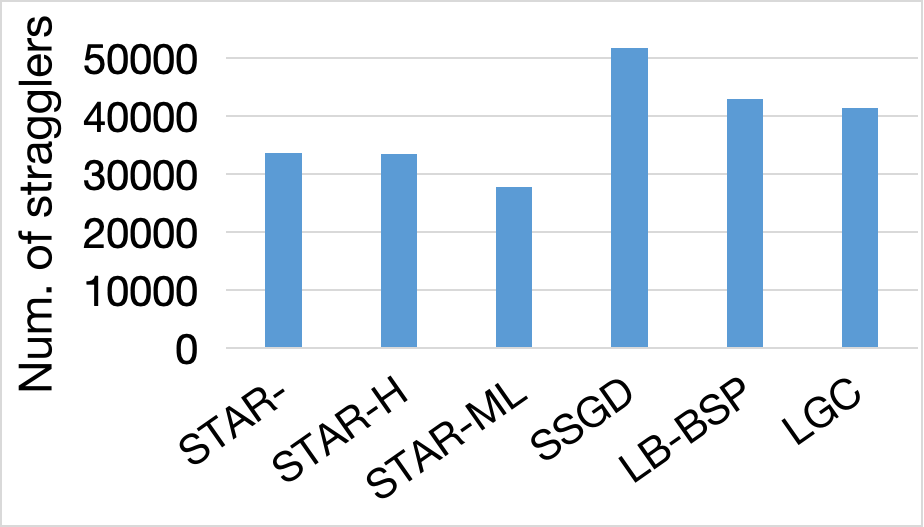}\label{subfig:concurr_strg_num_allred}}
\vspace{-0.2in}\caption{The number of stragglers.}\vspace{-0.1in}
\label{fig:concurr_strg_num}
\end{figure}

\noindent\textbf{Number of stragglers.} Fig.~\ref{fig:concurr_strg_num} shows straggler counts across systems. In PS, ASGD, Zeno++, Sync-Switch, and LGC have 26\%, 24.1\%, 12\%, and 9.3\% more stragglers than SSGD, as ASGD (used in Zeno++ and Sync-Switch) consumes more resources. LGC's partial gradient aggregation reduces this overhead. LB-BSP yields similar counts as SSGD since it only adjusts mini-batch sizes. STAR-H has 24.1\% fewer stragglers than SSGD due to its prediction and prevention mechanisms. STAR-ML further reduces stragglers by 9.7\% over STAR-H through lower CPU contention. STAR-H also has 1.7\% fewer stragglers than STAR-, as its training pauses reduce concurrent resource contention.\looseness=-1

For the AR architecture, LB-BSP shows a similar number of stragglers as SSGD. LGC and STAR-H reduce stragglers by 19\% and 39\%, respectively, by excluding slower workers and reconnecting them to high-bandwidth peers. However, because LGC removes only $N-K$ workers, some stragglers remain, making STAR-H more effective. STAR-ML has 15\% fewer stragglers and STAR- has 3\% more than STAR-H for the same reasons noted above.\looseness=-1


\vspace{-0in}
\subsection{Effectiveness of Individual Method}\vspace{-0in}

We tested variants of STAR-H and STAR-ML.
\squishlist
\vspace{-0.0in}    \item W/o our Straggler Prediction and instead uses the prediction method in~\cite{li2021sync} introduced in \cref{sec:exp_measure} (/SP).
    \vspace{-0.0in}    \item W/o dynamic-$x$ Synchronization or static-$x$ Synchronization and only has the ASGD option (/xS).
\vspace{-0.0in}    \item W/o Dynamic-$x$-order Synchronization (/DS).
\vspace{-0.0in}    \item W/o ``Preventing Stragglers upon mode change'' (/PS).
\vspace{-0.0in}    \item W/o Worker assignment in ``preventing stragglers upon mode change'' (/W).
\vspace{-0.0in}    \item W/o considering Resource sensitivity or training Stage in ``Preventing Stragglers upon mode change'' (/RS).
\vspace{-0.0in}    \item W/o Muri and choose the server that can host the most high-load tasks in ``high-load task assignment'' (/Mu).
\vspace{-0.0in}    \item Muri w/o balancing the number of high-load tasks (/N).\looseness=-1
\vspace{-0.0in}    \item W/o ``amortizing communication overhead'' (/Tree).
\squishend
For simplicity, we use STAR to represent either STAR-H or STAR-ML. We only include /SP, /DS, and /xS for STAR-H due to space limit.



\begin{figure}[h]\vspace{-0in}
\centering
\subfigure[PS architecture.]
{\includegraphics[width=0.49\columnwidth,height=2cm]{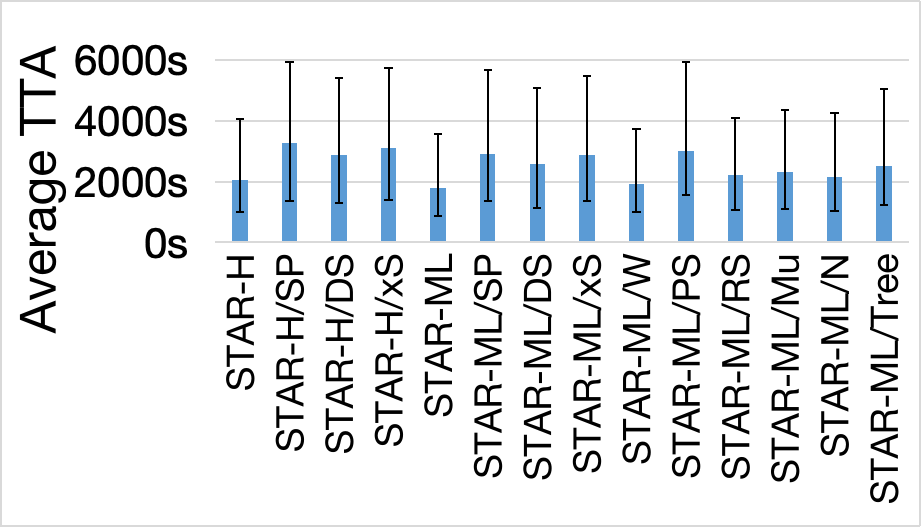}\label{subfig:concurr_avgtta_aws_indiv}}
\subfigure[All-reduce architecture. ]
{\includegraphics[width=0.49\columnwidth,height=2cm]{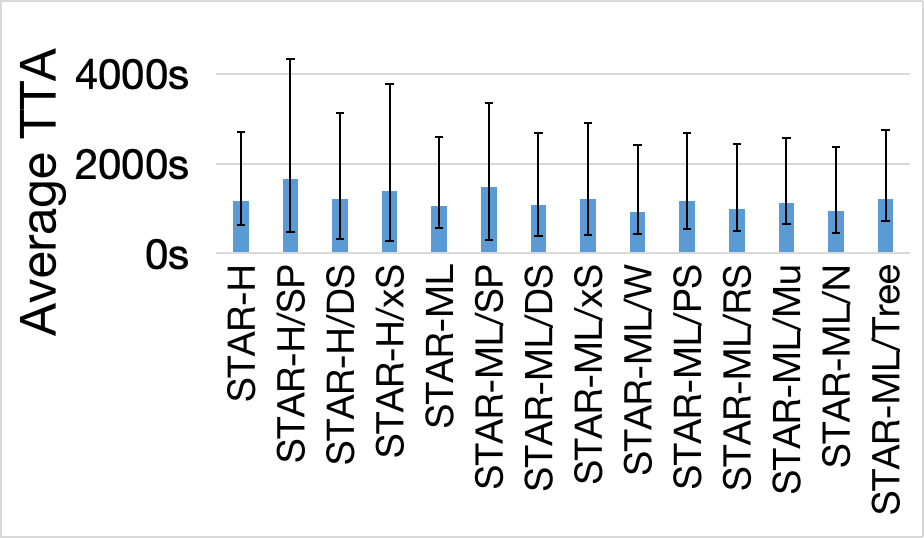}\label{subfig:concurr_avgtta_aws_indiv_allred}}
\vspace{-0.2in}\caption{TTA per job in STAR variants.}\vspace{-0.1in}
\label{fig:concurr_avgtta_indiv}
\end{figure}

\begin{figure}[h]\vspace{-0in}
\centering
\subfigure[PS architecture.]
{\includegraphics[width=0.49\columnwidth,height=2cm]{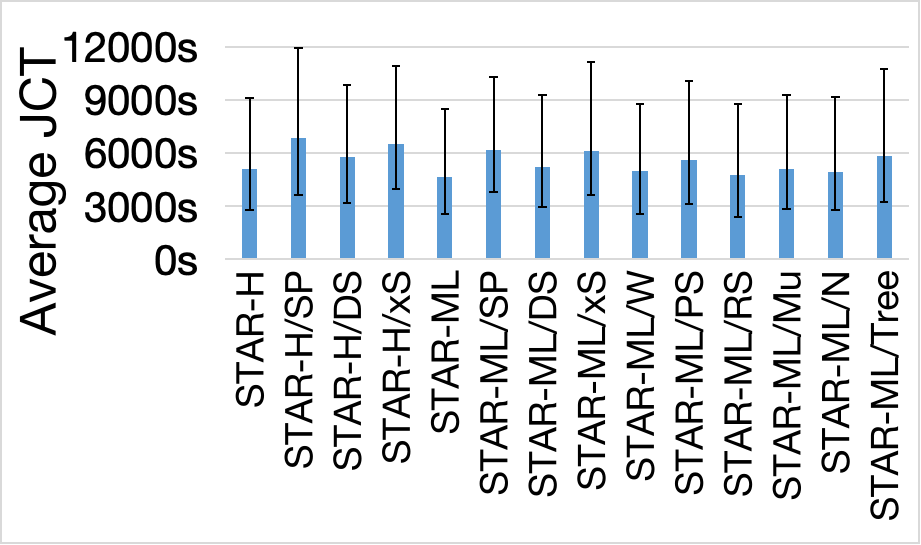}\label{subfig:concurr_avgjct_aws_indiv}}
\subfigure[All-reduce architecture.]
{\includegraphics[width=0.49\columnwidth,height=2cm]{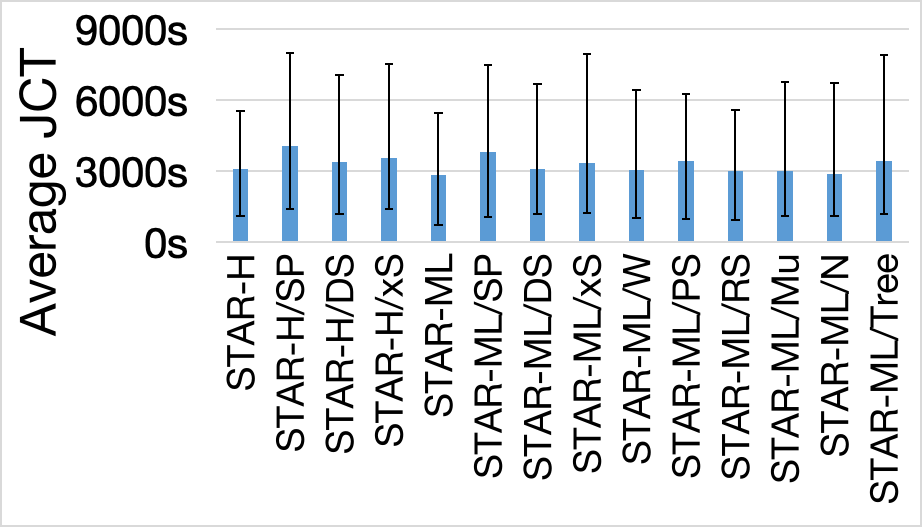}\label{subfig:concurr_avgjct_aws_indiv_allred}}
\vspace{-0.2in}\caption{JCT per job in STAR variants.}
\label{fig:concurr_avgjct_indiv}\vspace{-0.1in}
\end{figure}

\noindent\textbf{TTA and JCT.} Fig.~\ref{fig:concurr_avgtta_indiv} shows the average TTA per job and the 1st and 99th percentiles for STAR variants. In PS, /SP, /DS, and /xS incur 64-72\%, 47-50\%, and 59-74\% higher average TTA than STAR. /RS, /Mu, /N, and /Tree add 24\%, 31\%, 21\%, and 40\% overhead, respectively. Compared with STAR, /PS and /W yield 73\% and 6\% higher TTA, highlighting the benefits of STAR's integrated methods and factors. A similar trend appears in AR, confirming consistent effectiveness. Fig.~\ref{fig:concurr_avgjct_indiv} shows average JCT per job and its 1st and 99th percentiles, exhibiting the same relationships, which further validate each method's contribution to reducing delay.\looseness=-1

\begin{figure}[h]\vspace{-0.1in}
\centering
\subfigure[PS architecture.]
{\includegraphics[width=0.49\columnwidth,height=2cm]{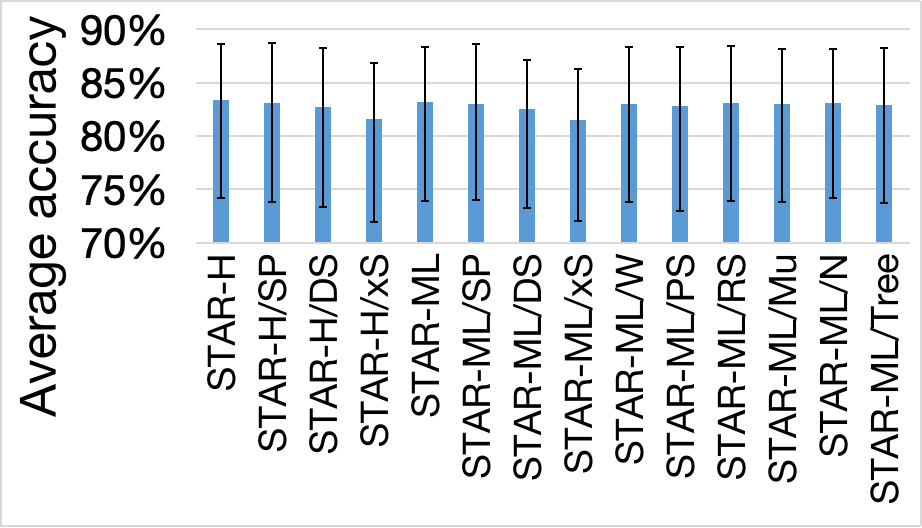}\label{subfig:concurr_avgacc_aws_indiv}}
\subfigure[All-reduce architecture.]
{\includegraphics[width=0.49\columnwidth,height=2cm]{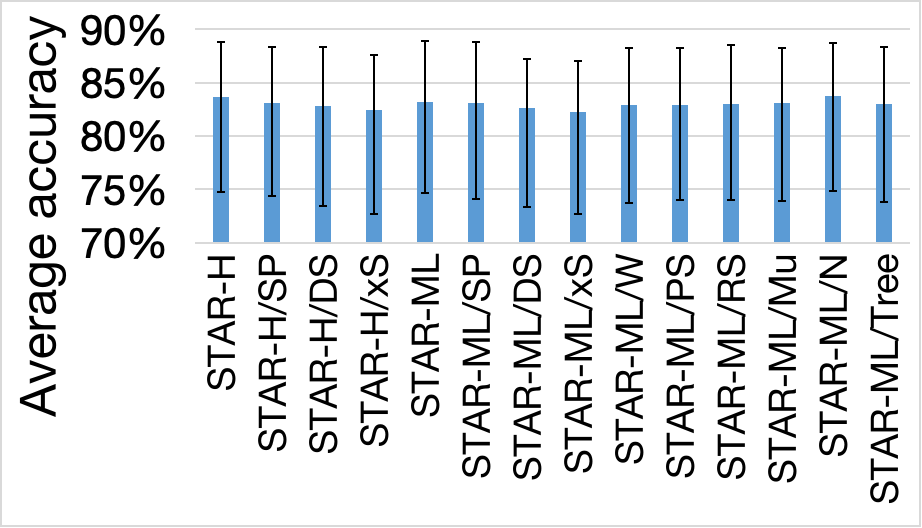}\label{subfig:concurr_avgacc_aws_indiv_allred}}
\vspace{-0.2in}\caption{Accuracy per image-classification job.}
\label{fig:concurr_avgacc_indiv}\vspace{-0.25in}
\end{figure}

\begin{figure}[h]\vspace{-0in}
\centering
\subfigure[PS architecture.]
{\includegraphics[width=0.49\columnwidth]{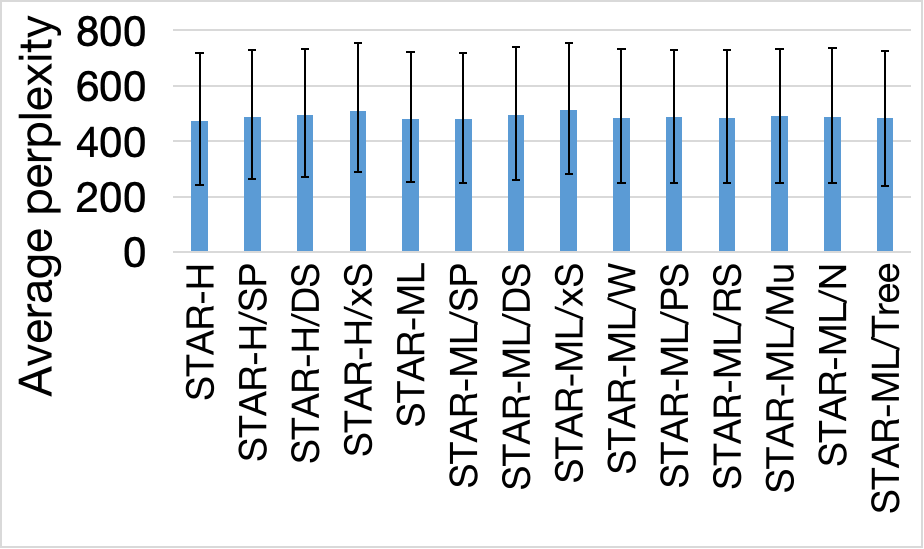}\label{subfig:concurr_avgperp_aws_indiv}}
\subfigure[All-reduce architecture.]
{\includegraphics[width=0.49\columnwidth]{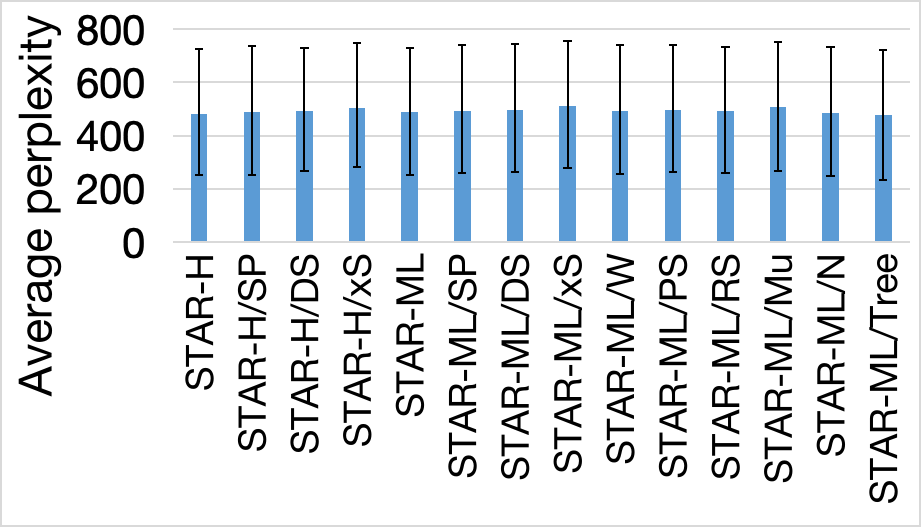}\label{subfig:concurr_avgperp_aws_indiv_allred}}
\vspace{-0.2in}\caption{Perplexity per NLP job.}
\label{fig:concurr_avgperp_indiv}\vspace{-0.1in}
\end{figure}

\noindent\textbf{Converged accuracy and perplexity.} Fig.~\ref{fig:concurr_avgacc_indiv} and Fig.~\ref{fig:concurr_avgperp_indiv} show the average converged accuracy and perplexity per job and their 1st and 99th percentiles for STAR variants. In PS, /SP shows 0.25-0.34\% lower accuracy and 0.39-1.26\% higher perplexity than STAR. /DS and /xS yield about 1.3\% and 2.5\% lower accuracy and 3.1\% and 7.3\% higher perplexity, respectively. /W has 0.16\% lower accuracy and 1.42\% higher perplexity. /PS, /RS, /Mu, /N, and /Tree show 0.1-0.6\% lower accuracy and 0.5-1.8\% higher perplexity. Similar patterns are seen in AR. These outcomes arise from failing to detect or prevent stragglers or from limited synchronization modes, which increase stale gradients and reduce accuracy, highlighting each method's effectiveness in mitigating such degradation.\looseness=-1

\begin{figure}[H]\vspace{-0in}
\centering
\subfigure[PS architecture.]
{\includegraphics[width=0.49\columnwidth,height=2cm]{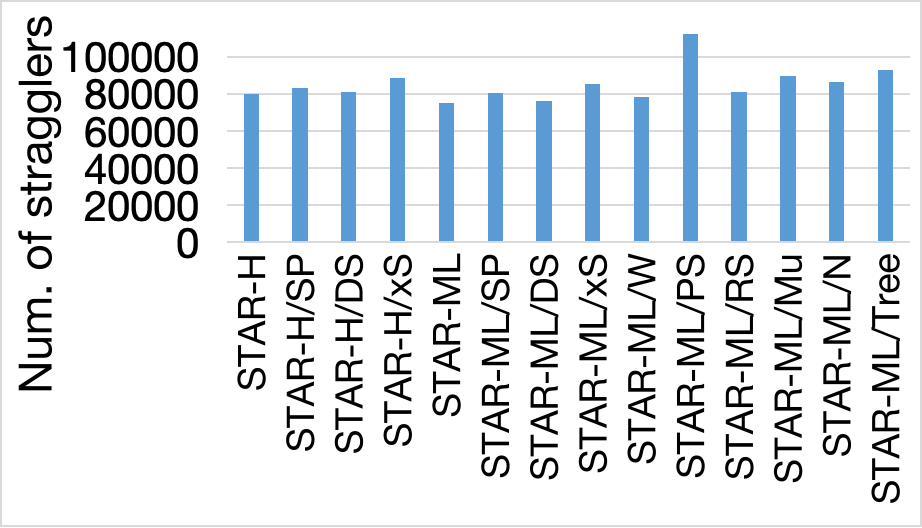}\label{subfig:concurr_strg_num_indiv_ps}}
\subfigure[All-reduce architecture.]
{\includegraphics[width=0.49\columnwidth,height=2cm]{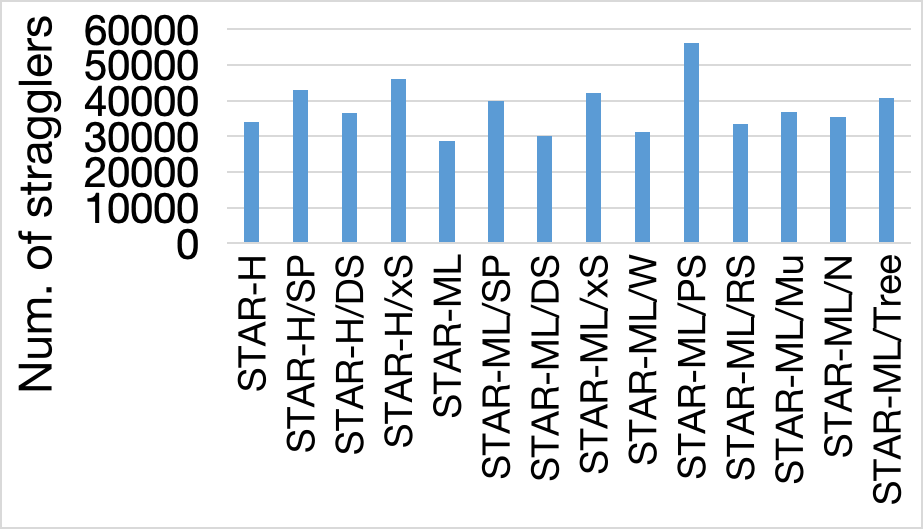}\label{subfig:concurr_strg_num_indiv_allred}}
\vspace{-0.2in}\caption{Num. of stragglers in STAR variants.}
\label{fig:concurr_strg_num_indiv}\vspace{-0.2in}
\end{figure}

\noindent\textbf{Number of Stragglers.} Fig.~\ref{fig:concurr_strg_num_indiv} shows the number of stragglers for STAR variants. In PS, /SP yields 4.9-8.3\% more stragglers than STAR due to less accurate identification. /DS produces a similar count since removing the dynamic-$x$-order mode barely changes resource use. /xS causes 11-15\% more stragglers as ASGD consumes more resources. /W, /PS, /RS, /Mu, /N, and /Tree create 4.8\%, 51\%, 10\%, 20\%, 19\%, and 23\% more stragglers than STAR because omitting their methods leaves insufficient resources for selected modes or co-located jobs. In AR, the same trend holds, confirming each method's role in reducing stragglers. Across variants, higher average metrics correlate with greater variance.\looseness=-1



\vspace{-0in}
\subsection{Time Overhead}\vspace{-0in}

\begin{figure}[h]\vspace{-0.1in}
    \centering
    \begin{minipage}[t]{0.235\textwidth}
  \centering
\includegraphics[width=1\columnwidth,height=2.3cm]{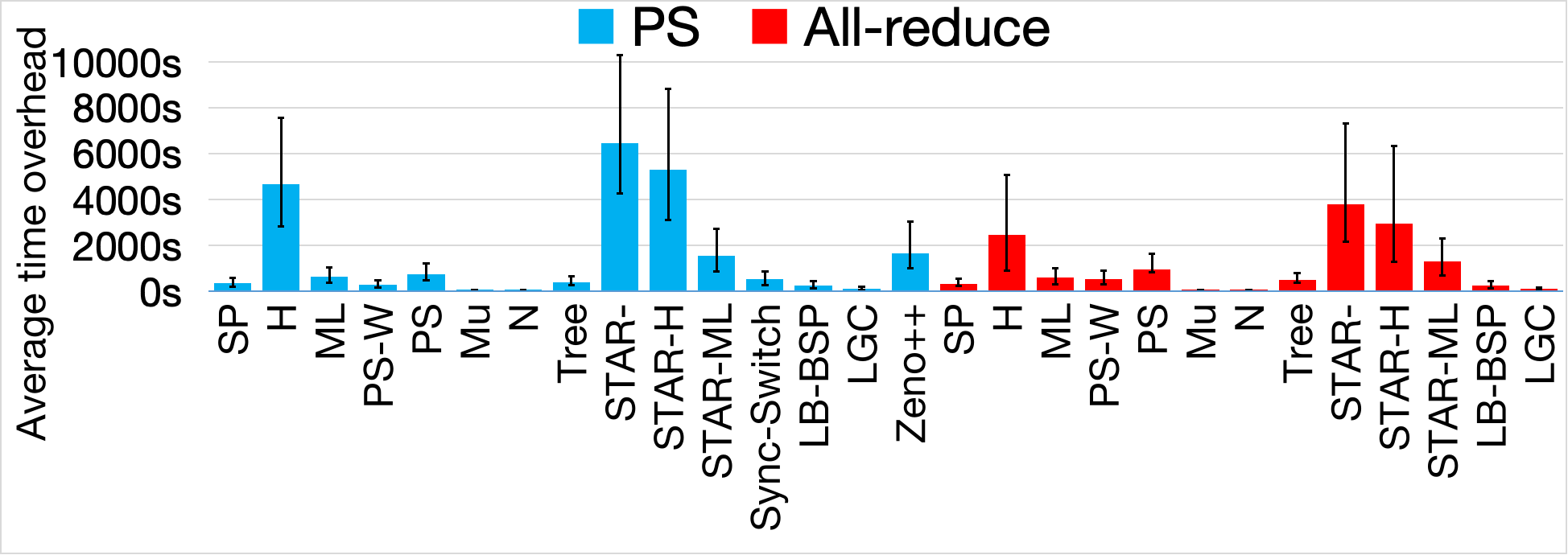}\label{subfig:concurr_time_overhead}
\vspace{-0.16in}\caption{Time overhead.}
\label{fig:concurr_time_overhead}\vspace{-0.2in}
    \end{minipage}%
\begin{minipage}[t]{0.235\textwidth}
  \centering
\includegraphics[width=1\columnwidth,height=2.3cm]{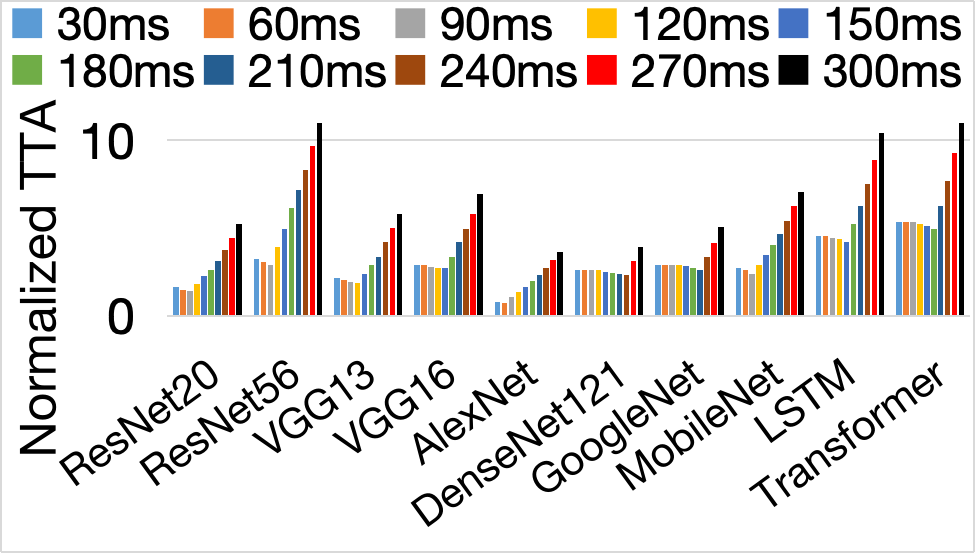}\label{subfig:allred_diff_delay}
\vspace{-0.16in}\caption{Different delays.}
\label{fig:allred_diff_delay}\vspace{-0.2in}
\end{minipage}\vspace{-0in}
\end{figure}

Fig.~\ref{fig:concurr_time_overhead} shows the average decision-making time per job and the 1st and 99th percentiles across systems and STAR variants. H and ML denote heuristic and ML inference, respectively. In both architectures, STAR-ML incurs about 200\%, 500\%, and 900\% higher overhead than Sync-Switch, LB-BSP, and LGC due to its additional components. Zeno++ has 8\% higher overhead than STAR-ML as it measures gradient staleness before applying gradients. However, since STAR-ML runs concurrently with training, it does not impact job progress, and its fewer stragglers (Fig.~\ref{fig:concurr_strg_num}) suggest its CPU use is non-intrusive. STAR-ML has 60-76\% lower average overhead than STAR-H, whose heuristic is slower. STAR- has 18-24\% higher overhead than STAR-H because it creates more stragglers and thus more decisions. ML accelerates H by 4.9-13x, and PS incurs 178\% more overhead than PS-W. Average overheads (in seconds) for H, PS, ML, Tree, SP, PS-W, Mu, and N are 4662, 736, 644, 385, 363, 282, 1.72, and 1.29 in PS, and 2583, 948, 595, 519, 481, 324, 1.32, and 1.15 in AR, respectively, reflecting their computational complexity. Generally, higher average overhead correlates with greater variance.

\subsection{Influence of Waiting Time in AR}\vspace{-0in}



Fig.~\ref{fig:allred_diff_delay} shows the normalized TTA for each job when AR parent workers wait 30-300ms for child gradients. For each model, TTA first decreases then increases as waiting time grows, with the minimum point being optimal and varying across models. Thus, the waiting time should be empirically determined for each model.\looseness=-1

\vspace{-0in}
\section{Related Work}\label{sec:related_work}\vspace{-0in}
Many methods leverage the asynchronous mode when stragglers appear~\cite{li2021sync,damaskinos2018asynchronous,xu2021lgc,zeno++, zhao2019dynamic,ho2013ssp,henggang2014boundedstaleness,chai2021fedat,lian2018asynchronous}. Sync-Switch~\cite{li2021sync} switches to ASGD when a straggler persists for a certain time period. Kardam \cite{damaskinos2018asynchronous} updates parameters using non-stale gradients from fast workers and either ignores or decays stale gradients from stragglers.
LGC \cite{xu2021lgc} uses the gradients from the $K$ fastest workers for the parameter update. Zeno++\cite{zeno++} is a variant of ASGD with bounded staleness, applying a small separate validation set to measure accuracy and updating parameters only when accuracy doesn't decrease. Dynamic Stale Synchronous Parallel (DSSP) \cite{zhao2019dynamic} dynamically adjusts the staleness threshold 
\cite{ho2013ssp, henggang2014boundedstaleness} to increase the frequency of fast workers' synchronous parameter updates. FedAT \cite{chai2021fedat} is a tier-based federated learning framework, which mitigates the straggler effect by updating local models synchronously within a tier (nodes with similar latency) and updating the global model asynchronously across tiers. AD-PSGD \cite{lian2018asynchronous} updates the model using an asynchronous and decentralized approach, in which each worker updates its local model with its neighbor. All the above methods neglect the possibility that the asynchronous mode may not improve TTA and could even create more stragglers due to its high resource consumption, observed from our measurement. STAR addresses these issues. 







Some approaches handle stragglers through worker duplication or replacement, or training data duplication~\cite{harlap2017proteus,luo2019hop,bitar2020stochastic,or2020resource}. Some work \cite{acar2013scheduling, harlap2016flexrr, zhou2019falcon, chen2019round, chen2020lbbsp, dean2012large} adjusts sizes of mini-batches of workers. A dynamic PS load distribution scheme (PSLD)~\cite{chen2020elastic} redistributes parameters among PSs to remove PS stragglers. SmartPS \cite{geng2019accelerating} prioritizes the transmission of updated parameters to stragglers. 
TicTac \cite{hashemi2019tictac} optimizes the order of parameter transmission to achieve near-optimal overlap between communication and computation. Unlike these approaches, STAR focuses on determining the optimal synchronization mode and can complement these methods in addressing stragglers.

\vspace{-0in}
\section{Limitations and Discussion}\label{sec:discussion}\vspace{-0in}
\noindent\textbf{PGNS.} PGNS requires significant time for each new model. We will explore faster ways to estimate training progress, such as tracking validation loss or analyzing gradient norms.

\noindent\textbf{Reduce time overhead.} STAR-ML maintains uninterrupted training by overlapping computation and training. To avoid pauses in STAR-H, future work will reduce its time overhead.

\noindent\textbf{Causes of stragglers.} Resource imbalance is a major cause of stragglers~\cite{abadi2016tensorflow}, but hardware, software, and network failures also contribute. We plan to extend STAR to handle these cases and evaluate it under such conditions.\looseness=-1

\noindent\textbf{Increase prediction accuracy.} Straggler prediction can be improved. We will investigate DL-based models that directly predict worker iteration times in one step.\looseness=-1


%


\vspace{-0.1in}

\section{Conclusion}\label{sec:conclusion}\vspace{-0in}

We studied straggler issues in homogeneous GPU-based DL training through extensive experiments and proposed STAR for PS and AR. STAR adds synchronization modes, selects the optimal one to minimize TTA, and ensures sufficient execution resources, with proactive straggler avoidance. Trace-driven AWS experiments show STAR reduces TTA by 48-84\% and 51-70\%, and JCT by 33-64\% and 55-77\% in PS and AR, respectively, achieving accuracy comparable to SSGD. No ethical issues arise, and future work will address remaining limitations.\looseness=-1

\bibliographystyle{unsrt}
\bibliography{reference}

@inproceedings{paszke2019pytorch,
  title     = {Pytorch: An imperative style, high-performance deep learning library},
  author    = {Paszke, Adam and Gross, Sam and Massa, Francisco and Lerer, Adam and Bradbury, James and Chanan, Gregory and Killeen, Trevor and Lin, Zeming and Gimelshein, Natalia and Antiga, Luca and others},
  booktitle = {Proc. of Advances in neural information processing systems},
  pages     = {8026--8037},
  year      = {2019}
}

@inproceedings{chen2015mxnet,
  author                      = {Tianqi Chen and Mu Li and Yutian Li and Min Lin and Naiyan Wang and Minjie Wang and Tianjun Xiao and Bing Xu and Chiyuan Zhang and Zheng Zhang},
  address                     = {Montréal, Canada},
  booktitle                   = {Proc. of Neural Information Processing Systems, Workshop on Machine Learning Systems},
  month                       = {Dec. 7-12},
  title                       = {{MXN}et: A flexible and efficient machine learning library for heterogeneous distributed systems},
  year                        = {2015},
  title_with_no_special_chars = {MXNet A Flexible and Efficient Machine Learning Library for Heterogeneous Distributed Systems}
}

@inproceedings{li2014ps,
  title     = {Scaling distributed machine learning with the parameter server},
  author    = {Li, Mu and Andersen, David G and Park, Jun Woo and Smola, Alexander J and Ahmed, Amr and Josifovski, Vanja and Long, James and Shekita, Eugene J and Su, Bor-Yiing},
  booktitle = {Proc. of the 11th {USENIX} Symposium on Operating Systems Design and Implementation ({OSDI} 14)},
  pages     = {583--598},
  year      = {2014}
}

@article{gerbessiotis1994direct,
  title     = {Direct bulk-synchronous parallel algorithms},
  author    = {Gerbessiotis, Alexandros V and Valiant, Leslie G},
  journal   = {Journal of parallel and distributed computing},
  volume    = {22},
  number    = {2},
  pages     = {251--267},
  year      = {1994},
  publisher = {Elsevier}
}

@inproceedings{he2016resnet,
  title     = {Deep residual learning for image recognition},
  author    = {He, Kaiming and Zhang, Xiangyu and Ren, Shaoqing and Sun, Jian},
  booktitle = {Proceedings of the IEEE conference on computer vision and pattern recognition},
  pages     = {770--778},
  year      = {2016}
}

@inproceedings{kenton2019bert,
  title     = {BERT: Pre-training of Deep Bidirectional Transformers for Language Understanding},
  author    = {Kenton, Jacob Devlin Ming-Wei Chang and Toutanova, Lee Kristina},
  booktitle = {Proceedings of NAACL-HLT},
  pages     = {4171--4186},
  year      = {2019}
}

@inproceedings{ho2013ssp,
  title     = {{More effective distributed ML via a stale synchronous parallel parameter server}},
  author    = {Ho, Qirong and Cipar, James and Cui, Henggang and Lee, Seunghak and Kim, Jin Kyu and Gibbons, Phillip B and Gibson, Garth A and Ganger, Greg and Xing, Eric P},
  booktitle = {Proc. of Advances in neural information processing systems},
  pages     = {1223--1231},
  year      = {2013}
}

@inproceedings{zhao2019dynamic,
  title        = {Dynamic stale synchronous parallel distributed training for deep learning},
  author       = {Zhao, Xing and An, Aijun and Liu, Junfeng and Chen, Bao Xin},
  booktitle    = {Proc. of IEEE 39th International Conference on Distributed Computing Systems (ICDCS)},
  pages        = {1507--1517},
  year         = {2019},
  organization = {IEEE}
}

@inproceedings{li2021sync,
  author    = {Li, Shijian and Mangoubi, Oren and Xu, Lijie and Guo, Tian},
  booktitle = {Proc. of IEEE 41th International Conference on Distributed Computing Systems (ICDCS)},
  title     = {Sync-Switch: Hybrid Parameter Synchronization for Distributed Deep Learning},
  year      = {2021}
}

@article{krizhevsky2009cifar,
  title     = {Learning multiple layers of features from tiny images},
  author    = {Krizhevsky, Alex and Hinton, Geoffrey and others},
  year      = {2009},
  journal = {Citeseer, Toronto, ON, Canada}
}

@inproceedings{aurick2021pollux,
  author    = {Aurick Qiao and Sang Keun Choe and Suhas Jayaram Subramanya and Willie Neiswanger and Qirong Ho and Hao Zhang and Gregory R. Ganger and Eric P. Xing},
  title     = {Pollux: Co-adaptive Cluster Scheduling for Goodput-Optimized Deep Learning},
  booktitle = {Proc. of the 15th {USENIX} Symposium on Operating Systems Design and Implementation ({OSDI} 21)},
  year      = {2021},
  isbn      = {978-1-939133-22-9},
  pages     = {1--18},
  url       = {https://www.usenix.org/conference/osdi21/presentation/qiao},
  publisher = {{USENIX} Association},
  month     = jul
}

@article{sam2018empiricalmodel,
  author     = {Sam McCandlish and
                Jared Kaplan and
                Dario Amodei and
                OpenAI Dota Team},
  title      = {An Empirical Model of Large-Batch Training},
  journal    = {CoRR},
  volume     = {abs/1812.06162},
  year       = {2018},
  url        = {http://arxiv.org/abs/1812.06162},
  eprinttype = {arXiv},
  eprint     = {1812.06162},
  bibsource  = {dblp computer science bibliography, https://dblp.org}
}

@inproceedings{chai2021fedat,
  author    = {Chai, Zheng and Chen, Yujing and Anwar, Ali and Zhao, Liang and Cheng, Yue and Rangwala, Huzefa},
  title     = {FedAT: A High-Performance and Communication-Efficient Federated Learning System with Asynchronous Tiers},
  year      = {2021},
  isbn      = {9781450384421},
  publisher = {Association for Computing Machinery},
  address   = {New York, NY, USA},
  url       = {https://doi.org/10.1145/3458817.3476211},
  doi       = {10.1145/3458817.3476211},
  booktitle = {Proceedings of the International Conference for High Performance Computing, Networking, Storage and Analysis},
  articleno = {60},
  numpages  = {16},
  location  = {St. Louis, Missouri},
  series    = {SC '21}
}

@inproceedings{chen2020elastic,
  title     = {Elastic parameter server load distribution in deep learning clusters},
  author    = {Chen, Yangrui and Peng, Yanghua and Bao, Yixin and Wu, Chuan and Zhu, Yibo and Guo, Chuanxiong},
  booktitle = {Proc. of SoCC},
  year      = {2020}
}

@inproceedings{qiao2018litz,
  author    = {Aurick Qiao and Abutalib Aghayev and Weiren Yu and Haoyang Chen and Qirong Ho and Garth A. Gibson and Eric P. Xing},
  title     = {Litz: Elastic Framework for High-Performance Distributed Machine Learning},
  booktitle = {Proc. of 2018 {USENIX} Annual Technical Conference ({USENIX} {ATC} 18)},
  year      = {2018},
  isbn      = {978-1-939133-01-4},
  address   = {Boston, MA},
  pages     = {631--644},
  url       = {https://www.usenix.org/conference/atc18/presentation/qiao},
  publisher = {{USENIX} Association},
  month     = jul
}

@inproceedings{acar2013scheduling,
  author    = {Acar, Umut A. and Chargueraud, Arthur and Rainey, Mike},
  title     = {Scheduling Parallel Programs by Work Stealing with Private Deques},
  year      = {2013},
  isbn      = {9781450319225},
  publisher = {Association for Computing Machinery},
  address   = {New York, NY, USA},
  volume    = {48},
  number    = {8},
  issn      = {0362-1340},
  url       = {https://doi.org/10.1145/2442516.2442538},
  doi       = {10.1145/2442516.2442538},
  booktitle = {Proceedings of the 18th ACM SIGPLAN Symposium on Principles and Practice of Parallel Programming},
  pages     = {219–228},
  numpages  = {10},
  location  = {Shenzhen, China},
  series    = {PPoPP '13},
  month     = feb
}

@inproceedings{zhou2019falcon,
  title        = {Falcon: Towards computation-parallel deep learning in heterogeneous parameter server},
  author       = {Zhou, Qihua and Wang, Kun and Guo, Song and Lu, Haodong and Li, Li and Guo, Minyi and Sun, Yanfei},
  booktitle    = {Proc. of the 2019 IEEE 39th International Conference on Distributed Computing Systems (ICDCS)},
  pages        = {196--206},
  year         = {2019},
  organization = {IEEE}
}

@inproceedings{chen2019round,
  title        = {Round-robin synchronization: Mitigating communication bottlenecks in parameter servers},
  author       = {Chen, Chen and Wang, Wei and Li, Bo},
  booktitle    = {Proc. of the IEEE INFOCOM},
  pages        = {532--540},
  year         = {2019},
  organization = {IEEE}
}

@inproceedings{or2020resource,
  title   = {Resource elasticity in distributed deep learning},
  author  = {Or, Andrew and Zhang, Haoyu and Freedman, Michael},
  booktitle = {Proceedings of Machine Learning and Systems},
  volume  = {2},
  pages   = {400--411},
  year    = {2020}
}

@inproceedings{lian2018asynchronous,
  title        = {Asynchronous decentralized parallel stochastic gradient descent},
  author       = {Lian, Xiangru and Zhang, Wei and Zhang, Ce and Liu, Ji},
  booktitle    = {Proc. of the International Conference on Machine Learning},
  pages        = {3043--3052},
  year         = {2018},
  organization = {PMLR}
}

@inproceedings{harlap2017proteus,
  author    = {Harlap, Aaron and Tumanov, Alexey and Chung, Andrew and Ganger, Gregory R. and Gibbons, Phillip B.},
  title     = {Proteus: Agile ML Elasticity through Tiered Reliability in Dynamic Resource Markets},
  year      = {2017},
  isbn      = {9781450349383},
  publisher = {Association for Computing Machinery},
  address   = {New York, NY, USA},
  url       = {https://doi.org/10.1145/3064176.3064182},
  doi       = {10.1145/3064176.3064182},
  booktitle = {Proceedings of the Twelfth European Conference on Computer Systems},
  pages     = {589–604},
  numpages  = {16},
  location  = {Belgrade, Serbia},
  series    = {EuroSys '17}
}

@inproceedings{damaskinos2018asynchronous,
  title        = {{Asynchronous Byzantine machine learning (the case of SGD)}},
  author       = {Damaskinos, Georgios and Guerraoui, Rachid and Patra, Rhicheek and Taziki, Mahsa},
  booktitle    = {Proc. of the International Conference on Machine Learning},
  pages        = {1145--1154},
  year         = {2018},
  organization = {PMLR}
}

@inproceedings{gu2019tiresias,
  title     = {Tiresias: A GPU cluster manager for distributed deep learning},
  author    = {Gu, Juncheng and Chowdhury, Mosharaf and Shin, Kang G and Zhu, Yibo and Jeon, Myeongjae and Qian, Junjie and Liu, Hongqiang and Guo, Chuanxiong},
  booktitle = {Proc. of NSDI},
  pages     = {485--500},
  year      = {2019}
}

@inproceedings{geng2019accelerating,
  title     = {Accelerating distributed machine learning by smart parameter server},
  author    = {Geng, Jinkun and Li, Dan and Wang, Shuai},
  booktitle = {Proceedings of the 3rd Asia-Pacific Workshop on Networking 2019},
  pages     = {92--98},
  year      = {2019}
}

@inproceedings{luo2019hop,
  title     = {Hop: Heterogeneity-aware decentralized training},
  author    = {Luo, Qinyi and Lin, Jinkun and Zhuo, Youwei and Qian, Xuehai},
  booktitle = {Proceedings of the Twenty-Fourth International Conference on Architectural Support for Programming Languages and Operating Systems},
  pages     = {893--907},
  year      = {2019}
}

@article{dean2012large,
  title={Large scale distributed deep networks},
  author={Dean, Jeffrey and Corrado, Greg and Monga, Rajat and Chen, Kai and Devin, Matthieu and Mao, Mark and Ranzato, Marc'aurelio and Senior, Andrew and Tucker, Paul and Yang, Ke and others},
  journal={Advances in neural information processing systems},
  volume={25},
  year={2012}
}

@inproceedings{henggang2014boundedstaleness,
  author    = {Henggang Cui and James Cipar and Qirong Ho and Jin Kyu Kim and Seunghak Lee and Abhimanu Kumar and Jinliang Wei and Wei Dai and Gregory R. Ganger and Phillip B. Gibbons and Garth A. Gibson and Eric P. Xing},
  title     = {Exploiting Bounded Staleness to Speed Up Big Data Analytics},
  booktitle = {Proc. of the USENIX Annual Technical Conference (USENIX ATC 14)},
  year      = {2014},
  isbn      = {978-1-931971-10-2},
  address   = {Philadelphia, PA},
  pages     = {37--48},
  url       = {https://www.usenix.org/conference/atc14/technical-sessions/presentation/cui},
  publisher = {USENIX Association},
  month     = jun
}

@inproceedings{chen2020lbbsp,
  author    = {Chen, Chen and Weng, Qizhen and Wang, Wei and Li, Baochun and Li, Bo},
  title     = {Semi-Dynamic Load Balancing: Efficient Distributed Learning in Non-Dedicated Environments},
  year      = {2020},
  isbn      = {9781450381376},
  publisher = {Association for Computing Machinery},
  address   = {New York, NY, USA},
  url       = {https://doi.org/10.1145/3419111.3421299},
  doi       = {10.1145/3419111.3421299},
  booktitle = {Proceedings of the 11th ACM Symposium on Cloud Computing},
  pages     = {431–446},
  numpages  = {16},
  location  = {Virtual Event, USA},
  series    = {SoCC '20}
}

@inproceedings{harlap2016flexrr,
  author    = {Harlap, Aaron and Cui, Henggang and Dai, Wei and Wei, Jinliang and Ganger, Gregory R. and Gibbons, Phillip B. and Gibson, Garth A. and Xing, Eric P.},
  title     = {{Addressing the Straggler Problem for Iterative Convergent Parallel ML}},
  year      = {2016},
  isbn      = {9781450345255},
  publisher = {Association for Computing Machinery},
  address   = {New York, NY, USA},
  url       = {https://doi.org/10.1145/2987550.2987554},
  doi       = {10.1145/2987550.2987554},
  pages     = {98--111},
  numpages  = {14},
  location  = {Santa Clara, CA, USA},
  booktitle    = {Proc. of SoCC '16}
}

@inproceedings{xu2021lgc,
  author    = {Xu, Jian and Huang, Shao-Lun and Song, Linqi and Lan, Tian},
  booktitle = {Proc. of IEEE INFOCOM 2021 - IEEE Conference on Computer Communications},
  title     = {Live Gradient Compensation for Evading Stragglers in Distributed Learning},
  year      = {2021},
  volume    = {},
  number    = {},
  pages     = {1-10},
  doi       = {10.1109/INFOCOM42981.2021.9488815}
}

@misc{phillytrace,
  title        = {{Microsfot Phyilly Trace.}},
  howpublished = {\url{https://github.com/msr-fiddle/philly-traces}},
  year = {2019},
}

@misc{alibaba-trace,
  title        = {{Alibaba PAI Production Cluster Data.}},
  howpublished = {\url{https://github.com/pcl-projects/Alibaba-PAI-Data/blob/main/StragglerMeasure.pdf}},
  year = {2022},
}

@misc{wikitext,
  title        = {The {WikiText} Long Term Dependency Language Modeling Dataset},
  howpublished = {\url{https://blog.salesforceairesearch.com/the-wikitext-long-term-dependency-language-modeling-dataset/}},
  year = {2016},
}

@inproceedings{hashemi2019tictac,
  title={Tictac: Accelerating distributed deep learning with communication scheduling},
  author={Hashemi, Sayed Hadi and Abdu Jyothi, Sangeetha and Campbell, Roy},
  booktitle={Proceedings of Machine Learning and Systems},
  volume={1},
  pages={418--430},
  year={2019}
}

@inproceedings{he2016deepresidual,
  author={He, Kaiming and Zhang, Xiangyu and Ren, Shaoqing and Sun, Jian},
  booktitle={Proc. of IEEE Conference on Computer Vision and Pattern Recognition (CVPR)},
  title={Deep Residual Learning for Image Recognition},
  year={2016},
  volume={},
  number={},
  pages={770-778},
  doi={10.1109/CVPR.2016.90}
}

@article{christopher2018measuring,
  author    = {Christopher J. Shallue and
               Jaehoon Lee and
               Joseph M. Antognini and
               Jascha Sohl{-}Dickstein and
               Roy Frostig and
               George E. Dahl},
  title     = {Measuring the Effects of Data Parallelism on Neural Network Training},
  journal   = {CoRR},
  volume    = {abs/1811.03600},
  year      = {2018},
  url       = {http://arxiv.org/abs/1811.03600},
  eprinttype = {arXiv},
  eprint    = {1811.03600},
  bibsource = {dblp computer science bibliography, https://dblp.org}
}

@article{priya2017accurate,
  author    = {Priya Goyal and
               Piotr Doll{\'{a}}r and
               Ross B. Girshick and
               Pieter Noordhuis and
               Lukasz Wesolowski and
               Aapo Kyrola and
               Andrew Tulloch and
               Yangqing Jia and
               Kaiming He},
  title     = {Accurate, Large Minibatch {SGD:} Training ImageNet in 1 Hour},
  journal   = {CoRR},
  volume    = {abs/1706.02677},
  year      = {2017},
  url       = {http://arxiv.org/abs/1706.02677},
  eprinttype = {arXiv},
  eprint    = {1706.02677},
  bibsource = {dblp computer science bibliography, https://dblp.org}
}

@inproceedings {abadi2016tensorflow,
author = {Mart{\'\i}n Abadi and Paul Barham and Jianmin Chen and Zhifeng Chen and Andy Davis and Jeffrey Dean and Matthieu Devin and Sanjay Ghemawat and Geoffrey Irving and Michael Isard and Manjunath Kudlur and Josh Levenberg and Rajat Monga and Sherry Moore and Derek G. Murray and Benoit Steiner and Paul Tucker and Vijay Vasudevan and Pete Warden and Martin Wicke and Yuan Yu and Xiaoqiang Zheng},
title = {{TensorFlow}: A System for {Large-Scale} Machine Learning},
booktitle = {12th USENIX Symposium on Operating Systems Design and Implementation (OSDI 16)},
year = {2016},
isbn = {978-1-931971-33-1},
address = {Savannah, GA},
pages = {265--283},
url = {https://www.usenix.org/conference/osdi16/technical-sessions/presentation/abadi},
publisher = {USENIX Association},
month = nov,
}

@inproceedings{zhao2022interleaving,
author = {Zhao, Yihao and Liu, Yuanqiang and Peng, Yanghua and Zhu, Yibo and Liu, Xuanzhe and Jin, Xin},
title = {Multi-Resource Interleaving for Deep Learning Training},
year = {2022},
isbn = {9781450394208},
publisher = {Association for Computing Machinery},
address = {New York, NY, USA},
url = {https://doi.org/10.1145/3544216.3544224},
doi = {10.1145/3544216.3544224},
booktitle = {Proceedings of the ACM SIGCOMM 2022 Conference},
pages = {428-440},
numpages = {13},
location = {Amsterdam, Netherlands},
series = {SIGCOMM'22}
}

@misc{ringallreduce,
  title        = {Bringing {HPC} Techniques to Deep Learning},
  howpublished = {\url{https://andrew.gibiansky.com/blog/machine-learning/baidu-allreduce/}},
  year = {2017},
}

@INPROCEEDINGS{oyama2016predicting,
  author={Oyama, Yosuke and Nomura, Akihiro and Sato, Ikuro and Nishimura, Hiroki and Tamatsu, Yukimasa and Matsuoka, Satoshi},
  booktitle={Proc. of 2016 IEEE International Conference on Big Data (Big Data)},
  title={Predicting statistics of asynchronous SGD parameters for a large-scale distributed deep learning system on GPU supercomputers},
  year={2016},
  volume={},
  number={},
  pages={66-75},
  doi={10.1109/BigData.2016.7840590}
}

@article{meng2019convergence,
title = {Convergence analysis of distributed stochastic gradient descent with shuffling},
journal = {Neurocomputing},
volume = {337},
pages = {46-57},
year = {2019},
issn = {0925-2312},
doi = {https://doi.org/10.1016/j.neucom.2019.01.037},
url = {https://www.sciencedirect.com/science/article/pii/S0925231219300578},
author = {Qi Meng and Wei Chen and Yue Wang and Zhi-Ming Ma and Tie-Yan Liu}
}

@inproceedings{zhang2016staleness,
author = {Zhang, Wei and Gupta, Suyog and Lian, Xiangru and Liu, Ji},
title = {Staleness-Aware Async-SGD for Distributed Deep Learning},
year = {2016},
isbn = {9781577357704},
publisher = {AAAI Press},
booktitle = {Proceedings of the Twenty-Fifth International Joint Conference on Artificial Intelligence},
pages = {2350-2356},
numpages = {7},
location = {New York, New York, USA},
series = {IJCAI'16}
}

@article{smith2018disciplined,
  doi = {10.48550/ARXIV.1803.09820},
  url = {https://arxiv.org/abs/1803.09820},
  author = {Smith, Leslie N.},
  title = {A disciplined approach to neural network hyper-parameters: Part 1 -- learning rate, batch size, momentum, and weight decay},
  publisher = {arXiv},
  year = {2018},
  journal={arXiv preprint arXiv:1803.09820},
  copyright = {arXiv.org perpetual, non-exclusive license}
}

@article{ranjan2016survey,
  title={A Survey on Techniques in NLP},
  author={Ranjan, Nihar and Mundada, Kaushal and Phaltane, Kunal and Ahmad, Saim},
  journal={International Journal of Computer Applications},
  volume={134},
  number={8},
  pages={6--9},
  year={2016},
  publisher={Foundation of Computer Science}
}

@inproceedings{ding2021decay,
author = {Ding, Yimin},
title = {The Impact of Learning Rate Decay and Periodical Learning Rate Restart on Artificial Neural Network},
year = {2021},
isbn = {9781450389273},
publisher = {Association for Computing Machinery},
address = {New York, NY, USA},
url = {https://doi.org/10.1145/3460268.3460270},
doi = {10.1145/3460268.3460270},
booktitle = {Proc. of the 2nd International Conference on Artificial Intelligence in Electronics Engineering},
pages = {6-14},
numpages = {9},
location = {Phuket, Thailand},
series = {AIEE 2021}
}

@ARTICLE{li2020packetio,
  author={Li, Xuesong and Cheng, Wenxue and Zhang, Tong and Ren, Fengyuan and Yang, Bailong},
  journal={IEEE Transactions on Parallel and Distributed Systems},
  title={Towards Power Efficient High Performance Packet I/O},
  year={2020},
  volume={31},
  number={4},
  pages={981-996},
  doi={10.1109/TPDS.2019.2957746}
}

@inproceedings{jiang2020unified,
author = {Yimin Jiang and Yibo Zhu and Chang Lan and Bairen Yi and Yong Cui and Chuanxiong Guo},
title = {A Unified Architecture for Accelerating Distributed {DNN} Training in Heterogeneous {GPU/CPU} Clusters},
booktitle = {14th USENIX Symposium on Operating Systems Design and Implementation (OSDI 20)},
year = {2020},
isbn = {978-1-939133-19-9},
pages = {463--479},
url = {https://www.usenix.org/conference/osdi20/presentation/jiang},
publisher = {USENIX Association},
month = nov,
}

@misc{adaptra,
      title={Adaptra: Straggler-Resilient Hybrid-Parallel Training with Pipeline Adaptation}, 
      author={Tianyuan Wu and Lunxi Cao and Hanfeng Lu and Xiaoxiao Jiang and Yinghao Yu and Siran Yang and Guodong Yang and Jiamang Wang and Lin Qu and Liping Zhang and Wei Wang},
      year={2025},
      eprint={2504.19232},
      archivePrefix={arXiv},
      primaryClass={cs.DC},
      url={https://arxiv.org/abs/2504.19232}, 
}

@article{malleus,
author = {Li, Haoyang and Fu, Fangcheng and Ge, Hao and Lin, Sheng and Wang, Xuanyu and Niu, Jiawen and Wang, Yujie and Zhang, Hailin and Nie, Xiaonan and Cui, Bin},
title = {Malleus: Straggler-Resilient Hybrid Parallel Training of Large-scale Models via Malleable Data and Model Parallelization},
year = {2025},
issue_date = {June 2025},
publisher = {Association for Computing Machinery},
address = {New York, NY, USA},
volume = {3},
number = {3},
url = {https://doi.org/10.1145/3725322},
doi = {10.1145/3725322},
journal = {Proc. ACM Manag. Data},
month = jun,
articleno = {185},
numpages = {28}
}

@inproceedings{megascale,
author = {Jiang, Ziheng and Lin, Haibin and Zhong, Yinmin and Huang, Qi and Chen, Yangrui and Zhang, Zhi and Peng, Yanghua and Li, Xiang and Xie, Cong and Nong, Shibiao and Jia, Yulu and He, Sun and Chen, Hongmin and Bai, Zhihao and Hou, Qi and Yan, Shipeng and Zhou, Ding and Sheng, Yiyao and Jiang, Zhuo and Xu, Haohan and Wei, Haoran and Zhang, Zhang and Nie, Pengfei and Zou, Leqi and Zhao, Sida and Xiang, Liang and Liu, Zherui and Li, Zhe and Jia, Xiaoying and Ye, Jianxi and Jin, Xin and Liu, Xin},
title = {MegaScale: scaling large language model training to more than 10,000 GPUs},
year = {2024},
isbn = {978-1-939133-39-7},
publisher = {USENIX Association},
address = {USA},
booktitle = {Proceedings of the 21st USENIX Symposium on Networked Systems Design and Implementation},
articleno = {41},
numpages = {16},
location = {Santa Clara, CA, USA},
series = {NSDI'24}
}

@inproceedings{understanding-straggler,
author = {Lin, Jinkun and Jiang, Ziheng and Song, Zuquan and Zhao, Sida and Yu, Menghan and Wang, Zhanghan and Wang, Chenyuan and Shi, Zuocheng and Shi, Xiang and Jia, Wei and Liu, Zherui and Wang, Shuguang and Lin, Haibin and Liu, Xin and Panda, Aurojit and Li, Jinyang},
title = {Understanding stragglers in large model training using what-if analysis},
year = {2025},
isbn = {978-1-939133-47-2},
publisher = {USENIX Association},
address = {USA},
booktitle = {Proceedings of the 19th USENIX Conference on Operating Systems Design and Implementation},
articleno = {27},
numpages = {16},
location = {Boston, MA, USA},
series = {OSDI '25}
}

@misc{chatgpt_statistics,
  howpublished = {\url{https://www.stylefactoryproductions.com/blog/chatgpt-statistics}},
  title = {{ChatGPT} Statistics - The Key Facts and Figures},
  year={2025}
}

@misc{openai_gpt3,
  howpublished = {\url{https://lambdalabs.com/blog/demystifying-gpt-3}},
  title = {{OpenAI's GPT-3} Language Model: A Technical Overview},
  year={2025}
}

@ARTICLE{gu2022liquid,
  author={Gu, Rong and Chen, Yuquan and Liu, Shuai and Dai, Haipeng and Chen, Guihai and Zhang, Kai and Che, Yang and Huang, Yihua},
  journal={IEEE Transactions on Parallel and Distributed Systems},
  title={Liquid: Intelligent Resource Estimation and Network-Efficient Scheduling for Deep Learning Jobs on Distributed GPU Clusters},
  year={2022},
  volume={33},
  number={11},
  pages={2808-2820},
  doi={10.1109/TPDS.2021.3138825}
}

@inproceedings{wang2020job,
author = {Wang, Haoyu and Liu, Zetian and Shen, Haiying},
title = {Job Scheduling for Large-Scale Machine Learning Clusters},
year = {2020},
isbn = {9781450379489},
publisher = {Association for Computing Machinery},
address = {New York, NY, USA},
url = {https://doi.org/10.1145/3386367.3432588},
doi = {10.1145/3386367.3432588},
booktitle = {Proceedings of the 16th International Conference on Emerging Networking EXperiments and Technologies},
pages = {108-120},
numpages = {13},
location = {Barcelona, Spain},
series = {CoNEXT '20}
}

@misc{ourcode,
  howpublished = {\url{https://anonymous.4open.science/r/STAR-Training-Sync-Mode}},
  title = {Source Code},
  year={2025}
}

@article{lstm,
    author = {Hochreiter, Sepp and Schmidhuber, Jurgen},
    title = "{Long Short-Term Memory}",
    journal = {Neural Computation},
    volume = {9},
    number = {8},
    pages = {1735-1780},
    year = {1997},
    month = {11},
    issn = {0899-7667},
    doi = {10.1162/neco.1997.9.8.1735},
    url = {https://doi.org/10.1162/neco.1997.9.8.1735},
    eprint = {https://direct.mit.edu/neco/article-pdf/9/8/1735/813796/neco.1997.9.8.1735.pdf},
}

@article{bitar2020stochastic,
	title={Stochastic gradient coding for straggler mitigation in distributed learning},
	author={Bitar, Rawad and Wootters, Mary and El Rouayheb, Salim},
	journal={IEEE Journal on Selected Areas in Information Theory},
	volume={1},
	number={1},
	pages={277--291},
	year={2020},
	publisher={IEEE}
}

@article{lee1985bin-packing,
author = {Lee, C. C. and Lee, D. T.},
title = {A Simple On-Line Bin-Packing Algorithm},
year = {1985},
issue_date = {July 1985},
publisher = {Association for Computing Machinery},
address = {New York, NY, USA},
volume = {32},
number = {3},
issn = {0004-5411},
url = {https://doi.org/10.1145/3828.3833},
doi = {10.1145/3828.3833},
journal = {J. ACM},
month = {jul},
pages = {562-572},
numpages = {11}
}

@misc{sync-swtich-code,
  howpublished = {\url{https://github.com/cake-lab/Sync-Switch}},
  title = {Sync-Switch's Code},
  year={2021}
}

@misc{agglomerative-clustering,
  howpublished = {\url{https://scikit-learn.org/stable/modules/generated/sklearn.cluster.AgglomerativeClustering.html}},
  title = {Scikit-learn: Agglomerative Clustering},
  year={2024}
}

@misc{psutil,
  howpublished = {\url{https://pypi.org/project/psutil/}},
  title = {Cross-platform lib for process and system monitoring in {Python}},
  year={2024}
}

@misc{cpulimit-linux,
  howpublished = {\url{https://manpages.ubuntu.com/manpages/xenial/man1/cpulimit.1.html}},
  title = {Limit the {CPU usage} of a process},
  year={2024}
}

@misc{tc-linux,
  howpublished = {\url{https://man7.org/linux/man-pages/man8/tc.8.html}},
  title = {Manipulate traffic control settings},
  year={2024}
}

@inproceedings{gao2020machine,
  title={Machine learning based workload prediction in cloud computing},
  author={Gao, Jiechao and Wang, Haoyu and Shen, Haiying},
  booktitle={2020 29th international conference on computer communications and networks (ICCCN)},
  pages={1--9},
  year={2020},
  organization={IEEE}
}

@InProceedings{zeno++,
  title = 	 {Zeno++: Robust Fully Asynchronous {SGD}},
  author =       {Xie, Cong and Koyejo, Sanmi and Gupta, Indranil},
  booktitle = 	 {Proceedings of the 37th International Conference on Machine Learning},
  pages = 	 {10495--10503},
  year = 	 {2020},
  editor = 	 {III, Hal Daume and Singh, Aarti},
  volume = 	 {119},
  series = 	 {Proceedings of Machine Learning Research},
  month = 	 {13--18 Jul},
  publisher =    {PMLR},
  url = 	 {https://proceedings.mlr.press/v119/xie20c.html},
}

@ARTICLE{time-varying,
  author={Guo, Chao and Li, Yongcheng and Yan, Yonghu and Chen, Wei and Bose, Sanjay Kumar and Shen, Gangxiang},
  journal={IEEE Transactions on Cloud Computing}, 
  title={Efficiently Consolidating Virtual Data Centers for Time-Varying Resource Demands}, 
  year={2022},
  volume={10},
  number={3},
  pages={1751-1764},
  doi={10.1109/TCC.2020.2997403}
}

\end{document}